\documentclass[aps, pra, superscriptaddress, twocolumn, preprintnumbers,amsmath,amssymb,notitlepage, nobalancelastpage,10pt,longbibliography]{revtex4-2}
\usepackage[colorlinks=true,citecolor=blue,linkcolor=blue,urlcolor=blue]{hyperref}
\usepackage{orcidlink}
\usepackage{amsmath}
\usepackage{verbatim}
\usepackage{graphicx}
\usepackage{color}
\usepackage{physics}
\usepackage{yfonts}
\usepackage{soul}
\usepackage[normalem]{ulem}
\usepackage{dsfont}
\usepackage{svg}

\newcommand{\mean}[1]{\left\langle{#1}\right\rangle}

\usepackage{color}
\newcommand{\stx}[1]{_\text{#1}}

\begin{document}

\title{Quantized Thouless pumps protected by interactions in dimerized Rydberg tweezer arrays}
\author{Sergi Juli\`a-Farr\'e\,\orcidlink{0000-0003-4034-5786}}
\email{sergi.julia-farre@pasqal.com}
\affiliation{ICFO - Institut de Ciencies Fotoniques, The Barcelona Institute of Science and Technology, Av. Carl Friedrich Gauss 3, 08860 Castelldefels (Barcelona), Spain}
\affiliation{PASQAL SAS, 7 rue Léonard de Vinci - 91300 Massy, Paris, France}
\author{Javier Argüello-Luengo\,\orcidlink{0000-0001-5627-8907}}
\affiliation{ICFO - Institut de Ciencies Fotoniques, The Barcelona Institute of Science and Technology, Av. Carl Friedrich Gauss 3, 08860 Castelldefels (Barcelona), Spain}
\affiliation{Departament de Física, Universitat Politècnica de Catalunya, Campus Nord B4-B5, 08034 Barcelona, Spain}
\author{Lo\"ic Henriet}
\affiliation{PASQAL SAS, 7 rue Léonard de Vinci - 91300 Massy, Paris, France}
\author{Alexandre Dauphin\,\orcidlink{0000-0003-4996-2561}}
\affiliation{PASQAL SAS, 7 rue Léonard de Vinci - 91300 Massy, Paris, France}

\begin{abstract}
We study Thouless pumps, i.e., adiabatic topological transport, in an interacting spin chain described by the dimerized \textit{XXZ} Hamiltonian. In the noninteracting case, quantized Thouless pumps can only occur when a topological singularity is encircled adiabatically. In contrast, here we show that, in the presence of interactions, such topological transport can even persist for exotic paths in which the system gets arbitrarily close to the noninteracting singularity. We illustrate the robustness of these exotic Thouless pumps through the behavior of the noninteracting singularity,  which for sufficiently strong interactions splits into two singularities separated by a spontaneous antiferromagnetic insulator. We perform a numerical benchmark of these phenomena by means of tensor network simulations of ground-state physics and real-time adiabatic dynamics. Finally, we propose an experimental protocol with Floquet-driven Rydberg tweezer arrays.
\end{abstract}

\maketitle
\section{Introduction}
In one-dimensional (1D) systems, nontrivial topological insulators can only be defined in the presence of at least one protecting symmetry, leading to the notion of 1D symmetry-protected phases (SPT)~\cite{Ryu2010_periodictable}. Interestingly, while at the ground-state level the presence of a symmetry is crucial to distinguish between trivial and topological ground states, the dynamical response of these systems to the breaking of such protecting symmetries can also be used to reveal their topological nature. This is the case of Thouless pumps~\cite{Thouless1983_prb,Nijs1989,AidelsburgerThouless2023}, in which the breaking of the chiral and inversion symmetry in a periodic parameter path leads to quantized charge transport. The latter can be understood as a dimensional reduction of the two-dimensional (2D) quantum Hall effects~\cite{Klitzing1980,TKNN1982}, by considering the time as a second dimension, and provides a direct measure of the topological invariant of the 1D SPT phase. 

In the last decade, we have witnessed the first experimental realizations of Thouless pumps in photonic platforms~\cite{TopologicalPhotonicsreview_2019,Cerjan2020Thouless,Jurguensen2021Thouless,Kraus2012_pump} as well as in cold atom quantum simulators~\cite{MaciejBook,Bloch_2008,Cooper_2019} with fermions~\cite{Nakajima2016,Minguzzi2022} and bosons~\cite{Lohse2016}, modeled by noninteracting Rice-Mele Hamiltonians~\cite{RiceMele1982}.  In the Rice-Mele model, the Thouless pump is implemented by modulating a dimerized single-particle hopping in a closed path that encloses the trivial-to-topological transition point. The simultaneous periodic modulation of a staggered on-site potential breaking the protecting chiral and inversion symmetry avoids the singularity at the transition point and the gap remains open during the cycle, leading to the quantized pump.

Thouless pumps also provide a natural framework to study the effect of many-body interactions in topological phases, a topic that has attracted great interest over the last years~\cite{rachel_2010}. In this direction, several works~\cite{Berg2011,Greschner2020,hayward2018,mondal2021_inheritance, Lin2020_dimerized,Mondal2022,Kuno2020,padhan2023interacting,athanasiou2023thouless} pointed towards situations where bosonic interactions do not destroy Thouless pumps, but rather lead to interaction-induced phenomena. For fermionic systems, the eventual breakdown of quantized Thouless pumps in the presence of onsite fermionic interactions was theoretically explained as a direct consequence of emergent many-body Mott physics~\cite{Nakagawa2018,Bertok2022}, and it has been recently observed in the experiment of Ref.~\cite{Walter2022}. However, Thouless pumps can also be observed in interacting fermionic systems due to competing effects~\cite{arguello2023stabilization,segura2023charge} in models hosting a spontaneous bond-order-wave phase ~\cite{Nakamura99, Nakamura2000,julia-farre2021revealing}. Finally, recent theory~\cite{Bertok2022} and experimental~\cite{viebahn2023interactioninduced} works have shown that, in the presence of finite-size gaps, quasiquantized fermionic Thouless pumps can be observed along displaced paths that do not encircle the noninteracting topological singularity.

\begin{figure}[b]
    \centering
\includegraphics[width=1\columnwidth]{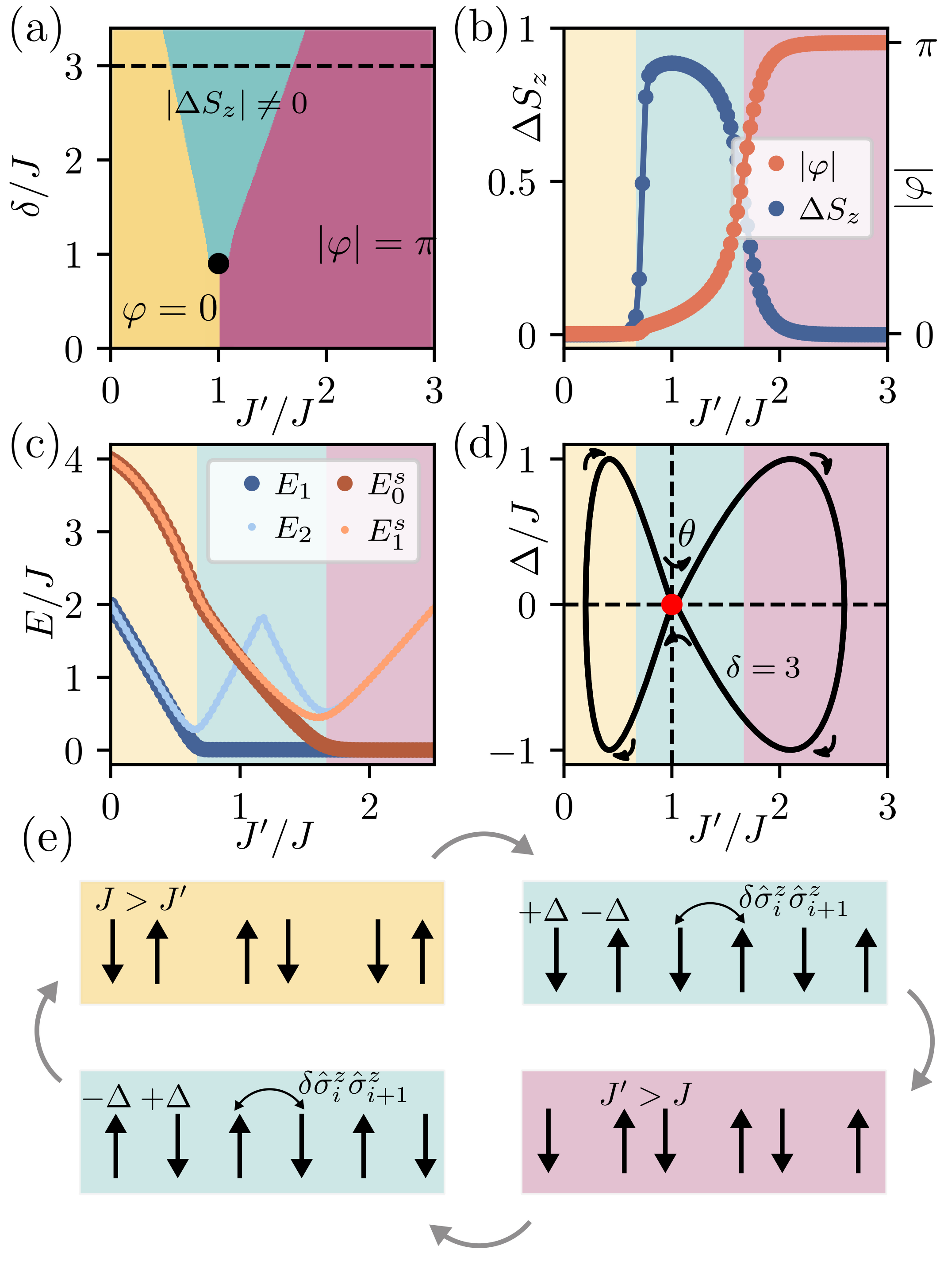}
    \caption{(a) Phase diagram of $H$ in the chiral-symmetric regime ($\Delta=0$). The tricritical point, indicated with a black dot, separates the three different phases: trivial dimerized (yellow), topologically dimerized (purple), and spontaneous antiferromagnetic phase (green). (b) Order parameters across the dashed line depicted in panel (a) at $\delta=3$. The trivial and topological chiral-symmetric dimerized phases are characterized by a quantized Zak phase $\varphi$, whereas the antiferromagnetic phase is signaled by a spontaneous finite local magnetization $\Delta S_z$. (c) Energy gaps along the same line at $\delta=3$. (d) Closed path used to perform a Thouless pump at $\delta=3$. Interactions allow one to explore the singularity during the pump (red point) without closing the bulk gap, due to the presence of a spontaneous antiferromagnetic phase. (e) Sketch of the instantaneous ground state during the path of panel (d).}
    \label{fig:phase_diagram}
\end{figure}

An open question is whether quantized Thouless pumps can be observed in interacting systems, in the extreme case where the periodic adiabatic path contains the noninteracting topological singularity. In this work, we show that such exotic Thouless pumps are indeed possible and that they can be stabilized even in the absence of finite-size gaps. To present our findings, we consider the dimerized spin-$1/2$ \textit{XXZ} chain~\cite{elben2020}, which we simulate with matrix-product-states~\cite{schollwock_density-matrix_2011}, and can be realized in tweezer arrays of Rydberg atoms~\cite{Saffman2010,henriet_quantum_2020,browaeys_many-body_2020,deLeseleuc2019,Scholl2022}. We start by using the density-matrix renormalization-group (DMRG) algorithm to revisit the interacting phase diagram of the system, which for sufficiently large interactions features a spontaneous antiferromagnetic order between its topological and trivial dimerized phases~\cite{elben2020}.  We then introduce a closed path in the Hamiltonian parameter space, in which both a staggered external field and the dimerized exchange couplings are varied, while the {exchange anisotropy remains fixed}, see Fig.~\ref{fig:phase_diagram}(d). Unlike in the paradigmatic Thouless pump, this path contains the noninteracting topological singularity, which is split into two singularities away from the origin in the presence of the intermediate antiferromagnetic phase.  Subsequently, we simulate the real-time dynamics of Thouless pumps along this newly introduced path with the infinite time-evolving block decimation (iTEBD) method. This allows us to certify that the system remains adiabatic, i.e., gapped, in the thermodynamic limit for sufficiently large interactions, and that a quantized topological transport occurs. As a remark, 
note that this exotic quantized transport protocol is in contrast with others considered~\cite{Bertok2022,viebahn2023interactioninduced} in similar split singularity scenarios, in which  
transport is quantized only in the presence of finite-size gaps. For completeness, we also show that nonadiabatic and nonquantized transport is recovered in our system for vanishing or small interactions, as expected from the gap closing at the singularity. Due to the adiabatic nature of Thouless pumps, we furthermore use the time-dependent variational principle (TDVP) in finite chains, to show that the same phenomena can be studied if, instead of considering real-time dynamics, one considers the preparation of independent ground states along the closed path. This is motivated by the experimental realization of this phenomenon, in which the adiabatic state preparation of several ground states could be easier to simulate than a dynamical change of the Hamiltonian parameters. In this line, we also propose an experimental setup based on Floquet engineering in Rydberg tweezer arrays~\cite{Scholl2022} to realize the interacting model under consideration and benchmark the adiabatic state preparation of its ground states, suitable to study the quantization of Thouless pumps. 

This article is organized as follows: In Section~\ref{sec:model}, we present the Hamiltonian model and revisit its phase diagram. Section~\ref{sec:thouless} contains the main results of this article. We start by introducing the adiabatic path, and subsequently study the quantization of Thouless pumps along it, both through real-time dynamics as well as by means of an adiabatic state preparation approach. Section~\ref{sec:exp} contains the proposal for the experimental implementation of the model with Rydberg arrays of optical tweezers. Finally, we conclude in Section~\ref{sec:conclusions}, where we also provide an outlook of further research directions in the context of interacting Thouless pumps.

\section{Model and phase diagram}\label{sec:model}
We consider a one-dimensional chain of $N$ spins, described by a dimerized \textit{XXZ} Hamiltonian~\cite{elben2020}
\begin{equation}\label{eq:xxz_dimerized}
\begin{split}
    H = &\frac{J}{2}\sum_{i=1}^{N/2} \left(\sigma_{2i-1}^x\sigma_{2i}^x+\sigma_{2i-1}^y\sigma_{2i}^y+\delta \sigma_{2i-1}^z\sigma_{2i}^z\right)\\
    &+\frac{J'}{2}\sum_{i=1}^{N/2-1} \left(\sigma_{2i}^x\sigma_{2i+1}^x+\sigma_{2i}^y\sigma_{2i+1}^y+\delta \sigma_{2i}^z\sigma_{2i+1}^z\right)\\
    &+\frac{\Delta}{2}\sum_i (-1)^i\sigma^z_i,
    \end{split}
\end{equation}
where $\sigma^\mu$ are the spin-$1/2$ Pauli matrices. Here $J,J'>0$ are the dimerized exchange couplings between nearest-neighbor (NN) spins, which break site-inversion symmetry, $\delta$ is the anisotropy parameter, and $\Delta$ is a staggered on-site field breaking bond-inversion and chiral symmetry. First, notice that for $\delta=0$ one recovers the Rice-Mele model, in which a Thouless pump is implemented by modulating the dimerization parameter $J'/J$ and the chiral symmetry breaking parameter $\Delta$ to encircle the singularity $(\Delta, J'/J)=(0,1)$.  For $\delta \neq 0$, the model becomes interacting, as the exchange anisotropy maps to a density-density coupling in the hardcore boson picture. The motivation to work with this model is twofold. On the one hand, it accounts for the minimal modification of the Rice-Mele model needed for the Thouless pumps proposed in this work. On the other hand, such a model can be realized with state-of-the-art programmable arrays of Rydberg atoms~\cite{Scholl2022}. The phase diagram at $\Delta=0$, obtained with the infinite DMRG method, is shown in Fig.~\ref{fig:phase_diagram}(a). 

For $\delta<1$, the system is either in a trivial or topological dimerized phase separated by the transition point at $J'/J=1$~\cite{elben2020}. In particular, we use the quantization of the Zak phase (see Appendix~\ref{app:zak}) to identify these two phases. Alternatively, for $\delta>1$ this singular transition point is replaced by an intermediate antiferromagnetic phase in a finite range of $J'/J$ values, in which the system exhibits a finite value in $\Delta S_z\equiv S^z_{i}-S^z_{i+1}\neq 0$, capturing the spontaneous breaking of the Ising symmetry. In Figs.~\ref{fig:phase_diagram}(b) and \ref{fig:phase_diagram}(c) we study the particular line $\delta=3$ in the phase diagram with the finite DMRG method ($N=64$) and open boundary conditions, to illustrate the behavior of the relevant order parameters in the different phases. Note that, due to the finite size and open boundary condition used in this method, there is a slight shift of the phase transition boundaries with respect to Fig.~\ref{fig:phase_diagram}(a). The use of the finite DMRG method also provides access to the first-excited states of the system. In particular, we compute the two excitation energies $E_1, E_2$ in the unpolarized ground-state sector ($S_{z, \text{total}}=0$), and also the first two excitation energies $E_0^s, E_1^s$ in the spin-flipped sector ($S_{z, \text{total}}=+1$). The ground-state energy is set to $E_0=0$.

Let us first focus on the trivial dimerized phase that appears at small $J'/J$. As shown in Fig.~\ref{fig:phase_diagram}(b), this phase is characterized by a quantized trivial Zak phase $\varphi=0$, and it is also a fully gapped phase, as shown in Fig.~\ref{fig:phase_diagram}(c). For increasing $J'/J$, the system eventually enters the antiferromagnetic phase, 
in which there is a spontaneous breaking of the chiral and bond-inversion symmetries driven by interactions, as captured by the finite value of $\Delta S_z\neq 0$. At the transition trivial-antiferromagnetic, the spin gap remains open, but there is an internal gap closing in the sector $S_{z, \text{total}}=0$ associated with the spontaneous antiferromagnetic transition. Inside the antiferromagnetic phase, there is a ground-state degeneracy ($E_1=E_0=0$) associated with the spontaneous symmetry breaking, but this manifold is gapped from the rest of the spectrum ($E_0^s,E_2 > 0$). Moreover, notice that the Zak phase is not quantized in the antiferromagnetic region, since 
its quantization relies on the presence of inversion symmetry, which is spontaneously broken in the Néel-like states characteristic of this phase. For even larger values $J'/J$, there is a transition to the topological dimerized phase. In this antiferromagnetic-topological phase transition, there is also a bulk gap closing in the sector $S_{z, \text{total}}=0$ associated with the spontaneous antiferromagnetic transition. Moreover, the spin gap also closes because the degenerate spin-polarized topological edge states emerge continuously from the gapped spin soliton band of the antiferromagnetic phase.
In the topological phase, both $E_1$ and $E_0^s$ coincide with $E_0$, due to the appearance of a degenerate edge-state manifold associated with the open boundary conditions, while $E_2$ and $E_1^s$ remain finite, showing that the bulk gap remains open in the topological phase.
The topological phase is characterized by a nonzero quantized Zak phase $\varphi=\pi$, as shown in Fig.~\ref{fig:phase_diagram}(b). 

In the context of Thouless pumps, the important conclusion from the phase diagrams of Figs.~\ref{fig:phase_diagram}(a)-\ref{fig:phase_diagram}(c), is that, in the presence of large interactions $\delta >1$, the singularity $J'/J=1$ splits into two singularities separated by an intermediate antiferromagnetic region. Furthermore, this antiferromagnetic region is effectively gapped, appart from the twofold degeneracy of the antiferromagnetic Néel states.

\section{Thouless pumps}\label{sec:thouless}
For a sufficiently large interaction, $\delta > 1$, a Thouless pump can be implemented for the Hamiltonian in Eq.~\eqref{eq:xxz_dimerized}, following the path illustrated in Fig.~\ref{fig:phase_diagram}(d), which constitutes the central result of this work. The path is parametrized by
\begin{equation}
\begin{split}
    J'(\theta)/J &=  1.4 + 1.2\cos(\theta)\\
    \Delta(\theta)/J &= 
    \begin{cases}
-\sin (a_1\theta)/2\, \hspace{5mm} (\theta <\pi/2)\\
-\sin(a_2\theta)/2\, \hspace{5mm} (\pi/2<\theta <3\pi/2)\\
+\sin(a_1\theta)/2\, \hspace{5mm} (3\pi/2<\theta <2\pi)\\
\end{cases},
    \end{split}
    \label{eq:pump_cycle}
    \end{equation}
where $\theta \in [0,2\pi)$. Here the coefficients $a_1\approx 0.698,\, a_2\approx 1.4370$ are fixed by the condition that the path includes the noninteracting singularity $(J'/J,\Delta)=(1,0)$. The presence of an interaction-induced antiferromagnetic energy gap allows the system to preserve adiabaticity even at the noninteracting topological singularity. As illustrated in Fig.~\ref{fig:phase_diagram}(e), which sketches the instantaneous state of the quantum system along the path, the latter can be understood from the fact that the Ising interaction favors the same local order as an staggered magnetic field. While, at the noninteracting singularity $(J'/J,\Delta)=(1,0)$ the ground state is always twofold degenerate, this does not break the adiabaticy of the Thouless pump, since the two antiferromagneyic (AFM) states are gaped from the rest of the bulk spectrum, as shown in Fig.~\ref{fig:phase_diagram}(c). As a remark, in this exotic scenario one should also take into account possible gap closings related to the interaction-induced antiferromagnetic transition, which can nevertheless we avoided by a suitable choice of path parameters. In particular, our results only apply to closed paths that have the same general properties as the one depicted in Fig.~\ref{fig:phase_diagram}(d). More precisely, the closed path should be such that (i) the singular line $J'/J=1$ is crossed twice, one from $J'/J=1^-$ and one from $J'/J=1^+$, (ii) the line $\Delta=0$ is also crossed twice, one from $\Delta=0^-$ and one from $\Delta=0^+$. (iii) The crossings of the line $\Delta=0$ cannot occur at the singular point $(J'/J,\Delta)=(1,0)$. Note however, that this condition still allows the path to contain the singular point $(J'/J,\Delta)=(1,0)$. (iv) the path avoids a direct topological phase transition between the trivial and topological dimerized phases by entering the spontaneous antiferromagnetic phase ($\delta >1$), and (v) $\Delta(\theta)$ is finite at the dimerized-antiferromagnet transition points, to avoid a spontaneous symmetry-breaking transition, which would imply the presence of a bulk gap closing. 

Here $\theta$ can be interpreted as the time $t$ in real adiabatic time dynamics with $\theta=2\pi t/T$, where $T$ is the duration of a periodic cycle. Alternatively, $\theta$ can also be understood as a label for independently prepared ground states. In both cases, as in any Thouless pump encircling a topological singularity in a finite system with boundaries, one expects the quantization of the spin transferred from one edge to the other of the chain without a net accumulation in the bulk. Without loss of generality, we thus consider the spin transfer from the left edge to the right edge
\begin{equation}
    \Delta Q^l_\theta \equiv Q^l_{\theta}-Q^l_{0}-\sum_{i\in \text{jumps}}\delta Q^l_{\theta_i}=-\Delta Q^r_\theta.
    \label{eq:deltaQ}
\end{equation}
Here $Q^l_\theta\equiv \frac{1}{2}\sum_{i\in \text{left edge}}{\sigma}^z_i$ is the integrated net spin on the left edge of the system for a given value of $\theta$, and $\delta Q^l_{\theta_i}$ is the value of any discontinuity (jump) in $Q^l_\theta$ during the pump, related to the crossing of two edge states inside the gap~\cite{Hatsugai_bulkedge_2016}. A remark is in order here concerning the definition of the sites belonging to the left edge. In a finite system with open boundary conditions, one needs to consider that the chain edge is composed of a few sites, due to the fact that topological edge states are not fully localized on the first (last) sites of the chain, but instead exhibit an exponential decay. Importantly, $\delta Q^l_{\theta_i}$ always vanishes if $\theta$ is interpreted as the time variable (i.e., the edge states are never pumped directly in real dynamics). However, if $\theta$ is interpreted as a label for independently prepared ground states, $\delta Q^l_{\theta_i}$ can be finite at values $\theta_i$ for which the edge-state manifold is degenerate.
The correspondence with the topological invariant is given by
\begin{equation}
    |\Delta Q^l_{2\pi}| = |\Delta \varphi|/\pi\in \mathbb{Z}_2,
    \label{eq:quantization_pump}
\end{equation}
where $|\Delta \varphi|$ is the change of the topological Zak phase across the topological singularity. 

\subsection{Real-time dynamics}\label{sec:realtime}
\begin{figure}[t]
    \centering
    \includegraphics[width=\columnwidth]{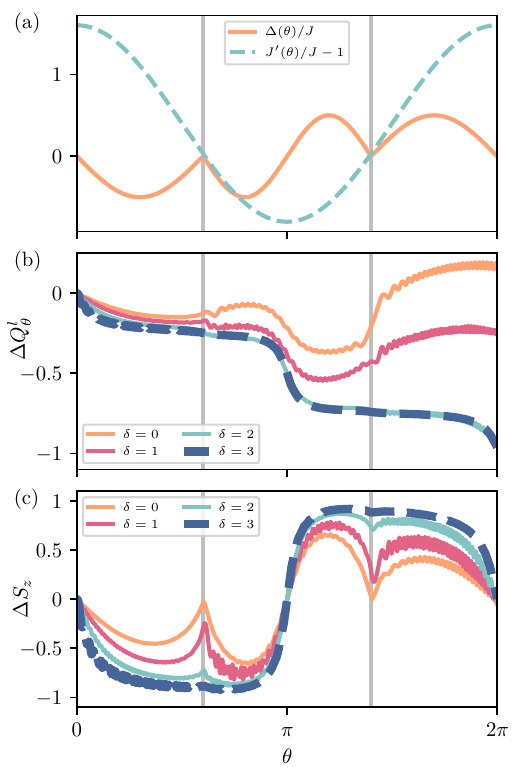}
    \caption{Real-time dynamics of Thouless pumps for different interaction strengths $\delta$, simulated using the iTEBD algorithm. (a) Modulation of the staggered detuning $\Delta$ and dimerization $J'$ during the Thouless pump along the path, given by Eqs.~\eqref{eq:pump_cycle} and sketched in Fig.~\ref{fig:phase_diagram}(d), with a total duration $TJ=200$. The values of $\theta$ in which the pump encounters the singularity $(\Delta, J'/J=0,1)$ are indicated with gray vertical lines. (b) Evolution of the transferred charge $Q^l_\theta$. (c) Evolution of the insulating antiferromagnetic order $\Delta S_z$.}
    \label{fig:thouless_TEBD}
\end{figure}

One option to experimentally realize Thouless pumps for the path illustrated in Fig.~\ref{fig:phase_diagram}(d) is to directly simulate the adiabatic dynamics, which requires the ability to dynamically modulate $J'(t)$ and $\Delta(t)$.  To investigate the quantization of a Thouless pump in this scenario, we simulate the real-time dynamics of such a protocol using the iTEBD method in an infinite chain (two-site unit cell). As an initial state, we consider the dimerized ground state of $H(\theta=0)$, obtained with infinite DMRG. Then, we simulate the adiabatic dynamics generated by $H(\theta)$, given by Eqs.~\eqref{eq:pump_cycle}, for a period of duration $TJ=200$, with a Trotter time-step of $\text{d}tJ=0.01$ in the iTEBD algorithm, and with a maximum bond dimension $\chi_\text{max}=200$. Note that even in a system without edges (an infinite or periodic chain), we can still compute the adiabatic spin transfer in the bulk by integrating the bulk spin current
\begin{equation}
\Delta Q^l_\theta = \int_0^{t(\theta)} \mathcal{J}(t')\, \text{dt}',
\end{equation}
where 
\begin{equation}
    \mathcal{J}(t)=\frac{i}{2}\sum_{j=j_\text{bulk}, j_\text{bulk}+1} J_j\mean{\sigma^+_{j}\sigma^-_{j+1}-\text{H.c.}}, \ \ J_j=J,J'. 
\end{equation}
Figure~\ref{fig:thouless_TEBD} shows the effect of interactions in Thouless pumps along the path of Fig.~\ref{fig:phase_diagram}(d). For completeness, in Fig.~\ref{fig:thouless_TEBD}(a) we show the numerical values of $J'(t)$ and $\Delta(t)$ as a function of the angle $\theta$. Remarkably, for sufficiently strong interactions, in Fig.~\ref{fig:thouless_TEBD}(b) we observe a perfectly quantized spin transport $|\Delta Q^l_{2\pi}|=1$ that, following Eq.~\eqref{eq:quantization_pump}, can be used to state that we have crossed a topological singularity adiabatically (i.e., without closing the bulk gap). The presence of a finite gap during the whole interacting pump can also be inferred from Fig.~\ref{fig:thouless_TEBD}(c), which shows the evolution of the antiferromagnetic order. Importantly, note that such antiferromagnetic insulating order remains robust except for the points $\theta=0, \pi, 2\pi$, where the gap of the system is dominated by the external dimerization $J'(\theta)/J\neq 1$. In particular, one can see that $\Delta S_z$ remains finite when the staggered magnetic field $\Delta$ vanishes (for the path of Eqs.~\eqref{eq:pump_cycle} this occurs approximately around $\theta \approx \pi/2,3\pi/2$), due to the presence of the spontaneous antiferromagnetic phase. In stark contrast with the interacting case, in the noninteracting or weakly interacting cases (orange and purple lines) we observe the expected breakdown of the Thouless pump, due to the fact that the periodic cycle encounters the singularity $(\Delta,J'/J)=(0,1)$. In this case,  $|\Delta Q^l_{2\pi}|$ is not quantized, and we observe strong oscillations in $\Delta Q^l_\theta$, as a consequence of the coupling to excited states in the absence of an energy gap. Figure~\ref{fig:thouless_TEBD}(c) provides further insight into such a breakdown of the Thouless pump. In the noninteracting case, one can clearly observe the decrease of the insulating antiferromagnetic order around $\theta\approx\pi/2,3\pi/2$, caused by the absence of a staggered magnetic field $\Delta$. In this region, one also has $J'/J \simeq 1$, so that both the antiferromagnetic and dimerization gaps vanish.

\subsection{Independent state-preparation protocol}
\label{sec:independetStatePrep}
Another strategy to experimentally observe the quantization given by Eq.~\eqref{eq:quantization_pump} is to adiabatically prepare independent ground states sketched in Fig.~\ref{fig:phase_diagram}(e) that appear across the path of Fig.~\ref{fig:phase_diagram}(d). That is, as we are interested in slow adiabatic dynamics, the instantaneous state of the system during the pump corresponds to the ground state of the instantaneous $H(\theta)$. However, note that when preparing independent ground states one has to take into account possible degeneracies in $H(\theta)$, which lead to an ambiguous connection between the instantaneous state during real-time dynamics and such ground states. On the one hand, one needs to consider topological degeneracies associated with a degenerate edge-state manifold. More precisely, while the edge-state degeneracy does not play a significant role in real-time dynamics, the instantaneous ground state of $H(\theta)$ in the topological sector is twofold degenerate at $\Delta=0$ with open boundary conditions, and it exhibits a discontinuous behavior in its net edge spin when going from $\Delta =0^-$ to $\Delta =0^+$ due to the lifting of the edge-state degeneracy. Such jumps in the edge spin occupation can be easily taken into account following Eq.~\eqref{eq:deltaQ}. On the other hand, in the interacting case $\delta > 1$, we have the abovementioned twofold degeneracy associated with the spontaneous antiferromagnetic phase at the singularity $(J'/J,\Delta)=(1,0)$. While, as shown in Fig.~\ref{fig:thouless_TEBD}, such a degeneracy does not impact the adiabatic real-time dynamics, if one aims at preparing the ground state in this regime, a random arbitrary choice of the initial state would lead to any of the two degenerate Néel states with equal probability. In this case, 
the twofold degeneracy of the ground state at the singularity $(\Delta, J'/J)=(0,1)$ can be effectively avoided with a suitable state-preparation protocol. We propose to initialize the state-preparation protocol at $\theta=0$ in one of the two fully polarized antiferromagnets
\begin{equation}
\label{eq:antiferromagnet}
\begin{split}
    \ket{\uparrow\downarrow}^{\otimes N/2}&\equiv \ket{\uparrow \downarrow \dots \uparrow \downarrow},\\
    \ket{\downarrow\uparrow}^{\otimes N/2}&\equiv \ket{\downarrow \uparrow \dots \downarrow \uparrow},
    \end{split}
\end{equation}
which can be obtained as the ground states of $H$ in Eq.~\eqref{eq:xxz_dimerized} for $\Delta\to \infty$ and $\Delta\to -\infty$, respectively. From such an initial state, one then adiabatically varies $\Delta:\pm \infty \to\pm \Delta(\theta)$. To avoid an arbitrary or random choice between the two possible initial states, we propose the following: (i) In general, one should choose the initial state as $\ket{\uparrow\downarrow}^{\otimes N/2}$ or $\ket{\downarrow\uparrow}^{\otimes N/2}$ for $\Delta (\theta) <0$ or $\Delta (\theta)>0$. (ii) In the special cases where $\Delta=0$, one chooses the initial state that would mimic the real-time adiabatic dynamics. At $\theta=0,2\pi$, this is given by the sign of $\Delta(0^+)$ and $\Delta(2\pi^-)$, respectively. At other singular points $\theta^*$ during the pump, the initial state must suddenly change if $\text{sign}[\Delta(\theta^{*-})]\neq \text{sign}[\Delta(\theta^{*+})]$ and $\Delta S_z(\theta^*)=0$.

\begin{figure}[h]
    \centering
    \includegraphics[width=\columnwidth]{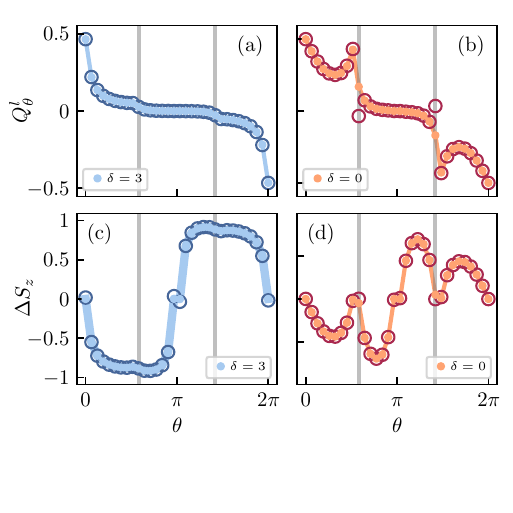}
    \caption{Results for independently prepared states across the path sketched in Fig.~\ref{fig:phase_diagram}(d), simulated using an adiabatic TDVP algorithm (see main text for details). We use a ramping time of $TJ=40$ for the state preparation.
    (a)-(b) Evolution of the left-edge net charge $Q^l_\theta$ for (a) $\delta=3$  and for (b) the noninteracting case.  (c)-(d) Evolution of the insulating antiferromagnetic order $\Delta S_z$. In all the panels, the empty circles show the values after letting the system evolve with the final Hamiltonian $H(\theta)$ during an extra time $\tau J=10$. }
    \label{fig:thouless_TDVP}
\end{figure}
We simulate such a protocol across the path of Fig.~\ref{fig:phase_diagram}(d) in a finite open chain of $N=64$ spins and consider a total duration of the ramping protocol of $TJ=40$, followed by a free evolution under $H(\theta)$ during an extra time $\tau J=10$, to ensure that the protocol leads to a stationary eigenstate of $H(\theta)$. The results for the left net spin $Q^l_\theta$ and insulating antiferromagnetic order $\Delta S_z$ are shown in Fig.~\ref{fig:thouless_TDVP}. A first observation is that by comparing these results with the real-time dynamics of Figs.~\ref{fig:thouless_TEBD}(b)-\ref{fig:thouless_TEBD}(c), there is a perfect agreement in the interacting case $\delta=3$ (blue points). The latter confirms that preparing independent ground states is a useful tool to mimic adiabatic dynamics in the system. Note that the main difference in this approach is that within such an independent state-preparation protocol, one, in general, would observe a quantized discontinuity in $Q^l_\theta$, which in Fig.~\ref{fig:thouless_TDVP} indeed occurs from $\theta=2\pi\rightarrow 0$. In the noninteracting case, the charge transfer is not quantized, as expected from the real-time dynamics results, but we observe important differences with respect to the iTEBD results. This is expected, as in the noninteracting case the population of excited states around the singularity $(J'/J,\Delta)=(1,0)$ leads to a breaking of the adiabatic nature of the dynamics, which cannot be captured by independently prepared ground states. Nevertheless, in Fig.~\ref{fig:thouless_TDVP}(b) we can also clearly see the breaking of adiabaticity in the noninteracting case, which shows a nonstationary value of $Q^l_\theta$ after the extra time evolution of $\tau J=10$. This highlights that one cannot adiabatically prepare the ground state when the gap of $H(\theta)$ vanishes.

\section{Implementation with programmable Rydberg arrays}\label{sec:exp}
\begin{figure}[t]
    \centering
\includegraphics[width=1.\columnwidth]{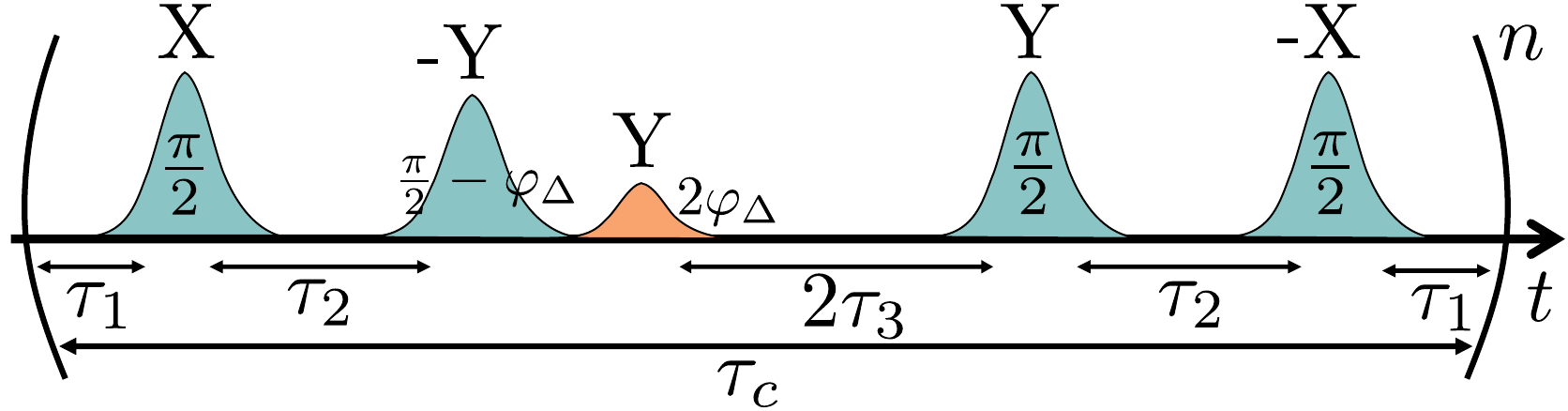}
    \caption{Pulse sequence associated with one Floquet cycle in the implementation of the \textit{XXZ} model in Eq.~\eqref{eq:xxz_dimerized}. Inside each pulse, we indicate its effective area (see text). Global pulses are shown in green, and those only applied on even sites, in red.}
    \label{fig:pulse_diagram}
\end{figure}
The dimerized \textit{XXZ} Hamiltonian in Eq.~\eqref{eq:xxz_dimerized} can be implemented with dipolar interactions. As a particular example, here we introduce a  scheme that can be directly implemented in current setups based on Rydberg arrays and optical tweezers~\cite{Nogrette2014,ebadiQuantum2022,barredoSynthetic2018,chenContinuous2023}. 
The internal structure of these atoms can be simplified as a two-level system with Rydberg states of opposite parity $\ket{nS}=\ket{\downarrow}$, and $\ket{nP}=\ket{\uparrow}$ coupled to each other through resonant dipole interactions of the form
\begin{equation}
H_{\textrm{XX}}= \sum_{i\neq j} \frac{J_{ij}}{2} (\sigma_i^x \sigma_j^x+\sigma_i^y \sigma_j^y)\,.
\end{equation}
The coupling strength, $J_{ij}=C_3 [1-3\cos^2(\theta_{ij})]/(2r_{ij}^3)$ depends on both the interatomic separation, $r_{ij}$, and the relative angle with the quantization axis, $\theta_{ij}$, which can be tuned by appropriately choosing the position of the optical tweezers.

These internal states can also be externally driven with a resonant microwave field of the form
\begin{equation}
H\stx{driven}= \hbar \Omega \sum_{i} (\cos(\phi) \sigma_i^x + \sin(\phi) \sigma_i^y),
\end{equation}
where $\Omega$ and $\phi$ are the amplitude and phase of the drive, which can be modified over time. An appropriate sequence of these pulses allows one to engineer the interactions of the system~\cite{geierFloquet2021b, vandersypenNMR2005a}.

For example a sequence of four global $\pi/2$ pulses of duration $\tau_p$ and constant phase $\phi=(0,-\pi/2,\pi/2,\pi)$ rotate the effective plane of the dipole interactions. This leads to a time-independent Hamiltonian of the form,
\begin{equation}
    H\stx{av}=\sum_{ij}\frac{J_{ij}}{\tau_c} [(\tau_1+\tau_2)\sigma_i^x \sigma_j^x+(\tau_1+\tau_3)\sigma_i^y \sigma_j^y + (\tau_2+\tau_3)\sigma_i^z \sigma_j^z]\,,
\end{equation}
where $\tau_{1,2,3}$ is the delay between pulses, and $\tau_c=2\sum_{i=1}^3\tau_i$ is the duration of the sequence, as it was experimentally implemented in Ref.~\cite{Scholl2022}. Using this procedure, one can access \textit{XXZ} models with $\delta\in [0,2]$.

In the following, we introduce in more detail how the independent state preparation approach described in Sec.~\ref{sec:independetStatePrep} applies to the simulation of the effective Hamiltonian in Eq.~\eqref{eq:xxz_dimerized}, which is engineered with the combination of the natural \textit{XX} dipole interaction and global and dimerized rotations along the $x$ and $y$-axis.
\begin{itemize}
    \item The system is formed by a dimerized chain of atoms placed in positions where the \textit{XX} interactions satisfy the desired ratio $J'/J$. For $J'\sim J$, this can correspond to a bipartite 1D chain with $r_2/r_1=(J'/J)^{1/3}$. Configurations where $J'/J\sim 0$ can also benefit from double ladders with neighbor atoms oriented along the magic angle $\theta_{ij}\sim 54.7 \, ^{\circ}$, where the dipole interactions vanish, as experimentally realized in Ref.~\cite{deLeseleuc2019}.
    \item  To prepare an antiferromagnetic state, one initializes all atoms in the state $\ket{\downarrow}$ using a stimulated Raman adiabatic passage. Next, a spatial light modulator (SLM) induces an Autler-Townes splitting of the atoms placed in odd positions, bringing them out-of-resonance from a last $\pi$ microwave rotation along the $y$-axis. Even if applied globally, the rotation only acts on the atoms that are masked by the SLM, providing experimental control of which atoms are flipped. This allows one to induce the fully polarized antiferromagnets of Eq.~\eqref{eq:antiferromagnet},

        \begin{equation}
\begin{split}
    \ket{\uparrow \downarrow }^{\otimes N/2}&\equiv \ket{\uparrow \downarrow  \dots \uparrow \downarrow }\,,\\
    \ket{\downarrow \uparrow }^{\otimes N/2}&\equiv \ket{\downarrow \uparrow  \dots \downarrow \uparrow }\,.
    \end{split}
\end{equation}

    \item To induce the \textit{XXZ} terms of the Hamiltonian, we perform a Floquet engineering of the effective dynamics following the procedure discussed above~\cite{geierFloquet2021b, vandersypenNMR2005a}. In particular, we choose $\tau_1=\tau_2$, and $\delta=(\tau_1+\tau_2)/(\tau_1+\tau_3)$. 
    
    \item The final step consists of inducing a staggered driving of the form $H_z=\Delta/2 \sum_i (-1)^i\sigma_i^z$. After the first $\pi/2$ $x$-rotation of the Floquet pulse, one can observe that, the operator $\sigma^z$ is mapped to $\sigma^y$. One can then induce $H_z$ by appropriately modifying the subsequent $(-Y)$-rotation of the Floquet sequence, which should be different on even and odd sites. In particular, one can apply a corrected effective area, $\pi/2-\varphi_\Delta$ on the global $(-Y)$-rotation, where $\varphi_\Delta= \tau_c \, \Delta/(1+\delta/2)$  (see the scheme in Fig.~\ref{fig:pulse_diagram}). Immediately after, the SLM is used to continue the rotation, on the even sites of the array, inducing a site-dependent short Gaussian $Y$-pulse of duration $\tau_p$ and area $2\varphi_\Delta$ that only affects those positions. We highlight that the pattern of the SLM remains constant throughout the implementation, preventing any limitation coming from its reconfiguration time.  Furthermore, the strength of $\Delta$ can be tuned over time by appropriately choosing its value on each cycle of the sequence.
    \item For the measurement stage of the protocol, a combination of fluorescence imaging and microwave rotations is conventionally used to measure the internal state of each atom~\cite{chenContinuous2023}.
    \end{itemize}

This protocol results in the implementation of the effective Hamiltonian in Eq.~\eqref{eq:xxz_dimerized} provided that the duration of each pulse is much shorter than $\tau_c$, and that this is shorter than the average interactions timescale, $N/\sum_{ij}J_{ij}$.
Let us now numerically benchmark this implementation for relevant parameters corresponding to a 1D tweezer array of $^{87}$Rb atoms with pseudospin states $\ket{\downarrow}=\ket{75S_{1/2},m_J=1/2}$ and $\ket{\uparrow}=\ket{75P_{3/2},m_J=3/2}$, which are separated by $2\pi\times 8.5$ GHz~\cite{Scholl2022}. These atoms are alternatively spaced with distances $r_{12}=19\,\mu$m, and $r_{23}=23.9\,\mu$m, so that the associated interaction is $J_{12}\approx 2\pi\times 270$ kHz, and $J'/J=2$. Site-dependent microwave pulses can be applied by using the SLM to induce a light shift on the $6P-75S$ transition on the undesired atoms, where laser beams of waist $\sim 2\,\mu$m
are achievable.

\begin{figure}[tbp]
    \centering
\includegraphics[width=\columnwidth]{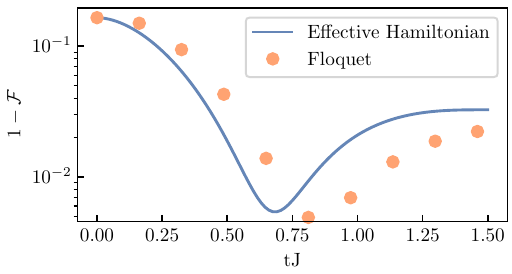}
    \caption{Infidelity in the preparation of state ($J'=2J$, $\delta=2J$, $\Delta=J$), $\mathcal{F}=\sqrt{\braket{\psi_\Delta}{\psi(t)}}$ when an initial AFM state in the $\sigma^x$-basis is evolved along the effective Hamiltonian in Eq.~\eqref{eq:xxz_dimerized} (blue line), or after each cycle of the Floquet protocol shown in Fig.~\ref{fig:pulse_diagram} (orange). Here we consider a 10-cycle preparation sequence of a chain with 12 atoms, and a cycle duration of $\tau_c=210$ ns.}
    \label{fig:fidelity}
\end{figure}

In Fig.~\ref{fig:fidelity}, we calculate the infidelity due to the state preparation ($J'=2J$, $\delta$=2, $\Delta=J$) when an initial AF state in the $x$-basis is evolved along the effective Hamiltonian in Eq.~\eqref{eq:xxz_dimerized} (blue line), or the Floquet protocol described above (orange). There, we consider a cycle duration of $\tau_c=210$ ns, Gaussian pulses with a $1/e^2$ width of $16$ ns and $J=1.7$ MHz. Following the procedure described above, we monotonically modify the effective area of the intermediate site-dependent $X$ pulses according to $\varphi_{\Delta\stx{F}(t)}$, with $\Delta\stx{F}(t)=\Delta+9\Delta\left[(T-t)/T \right] ^{1/4}$. For a conservative sequence of 10 cycles, we obtain a preparation infidelity of order $0.6\%$.

As a final remark, we note that, despite with Rydberg atoms in optical tweezers one does not typically have dynamical control over $J'/J$, the real-time adiabatic dynamics described in Sec.~\ref{sec:realtime} could be simulated for example with ultracold atoms in optical lattices. In that platform, the use of an optical superlattice allows one to simultaneously modulate in time the tunneling ratio $J'/J$ and the on-site field $\Delta$, which have already allowed to study Thouless pumps in the noninteracting hardcore boson limit~\cite{Lohse2016}, as well as with on-site interactions in fermionic systems~\cite{Walter2022}.  
While realizing the model of Eq.\eqref{eq:xxz_dimerized} requires engineering intersite interactions, these can be achieved, e.g., with Rydberg-dressing~\cite{weckesser2024realization}.

\section{Conclusions and outlook}\label{sec:conclusions}
In this work, we analyzed a class of Thouless pumps in interacting spin systems. In particular, we showed that, for sufficiently large interactions, the dimerized \textit{XXZ} chain exhibits quantized topological transport along a newly introduced path that contains the topological singularity, and that such an effect is completely absent for vanishing or small interactions. We could understand this phenomenon by the fact that interactions replace the original singularity by an intermediate antiferromagnetic phase that, despite being topologically trivial, has a finite bulk gap that can be used to preserve the adiabaticity of the cycle. Consequently, we expect that similar phenomena could also be observed in other models in which a direct topological phase transition is replaced by an intermediate spontaneous symmetry-breaking phase induced by interactions~\cite{Marks2021}. There, we stress that one should always ensure that the adiabatic path is such that it avoids bulk gap closings associated with spontaneous symmetry-breaking transitions. 
Finally, our experimental proposal paves the way for the experimental realization of the phenomena studied in this work using Rydberg tweezer arrays~\cite{henriet_quantum_2020,browaeys_many-body_2020}, as well as other emergent platforms also providing extended dipolar interactions~\cite{cornish2024quantum}.

\textit{Note added:} We became aware of the recent work~\cite{PhysRevB.109.235143}, in which the
authors study Thouless pumps for similar paths as the one
considered in the present article.

\acknowledgements
The tensor-network calculations were performed using the TeNPy Library~\cite{tenpy}. The ICFO group acknowledges support from ERC AdG NOQIA; MCIN / AEI(PGC20180910.13039 / 501100011033, CEX2019-000910S / 10.13039 / 501100011033, Plan National FIDEUA PID2019-106901GB-I00, Plan National STAMEENA PID2022-139099NB-I00 project funded by MCIN/AEI/10.13039/501100011033 and by the “European Union NextGenerationEU/PRTR" (PRTR-C17.I1), FPI); QUANTERA MAQS PCI2019-1118282); QUANTERA DYNAMITE PCI2022-132919 (QuantERA II Programme co-funded by European Union’s Horizon 2020 program under Grant Agreement No 101017733), Ministry of Economic Affairs and Digital Transformation of the Spanish Government through the QUANTUM ENIA project call – Quantum Spain project, and by the European Union through the Recovery, Transformation, and Resilience Plan – NextGenerationEU within the framework of the Digital Spain 2026 Agenda; Fundació Cellex; Fundació Mir-Puig; Generalitat de Catalunya (European Social Fund FEDER and CERCA program, AGAUR Grant No. 2021 SGR 01452, QuantumCAT \ U16-011424, co-funded by ERDF Operational Program of Catalonia 2014-2020); Barcelona Supercomputing Center MareNostrum (FI-2023-1-0013); EU Quantum Flagship (PASQuanS2.1, 101113690); EU Horizon 2020 FET-OPEN OPTOlogic (Grant No 899794); EU Horizon Europe Program (Grant Agreement 101080086 — NeQST), ICFO Internal “QuantumGaudi” project; European Union’s Horizon 2020 program under the Marie Sklodowska-Curie grant agreement No 847648;  “La Caixa” Junior Leaders fellowships, La Caixa” Foundation (ID 100010434): CF/BQ/PR23/11980043. Views and opinions expressed are, however, those of the author(s) only and do not necessarily reflect those of the European Union, European Commission, European Climate, Infrastructure and Environment Executive Agency (CINEA), or any other granting authority.  Neither the European Union nor any granting authority can be held responsible for them.  AD acknowledges funding from the European Union under Grant Agreement 101080142 and the project EQUALITY.

\appendix

\section{calculation of the Zak phase based on the entanglement spectrum}\label{app:zak}

We can use the relation between entanglement and quantized responses~\cite{Zaletel2014}, to compute the bulk Zak phase $\varphi$ in the matrix-product-state representation of the quantum state, which provides a direct access to the symmetry-resolved entanglement spectrum of a system. In particular, we use the relation~\cite{Zaletel2014}
\begin{equation}\label{eq:ent_spectrum}
e^{i \varphi} = \exp\left(2\pi i \sum_ps^2_pQ_p\right),
\end{equation}
where $s_p$ are the eigenvalues of the symmetry-resolved entanglement spectrum corresponding to the quantum number $Q_p$ of the conserved spin charge operator $\hat{N}\equiv \frac{1}{2}\sum_i (\hat{\sigma}^z_i+1)$. Note that, in a finite system, we compute the entanglement spectrum for a bipartition at the center of the chain.


\begin{thebibliography}{56}%
\makeatletter
\providecommand \@ifxundefined [1]{%
 \@ifx{#1\undefined}
}%
\providecommand \@ifnum [1]{%
 \ifnum #1\expandafter \@firstoftwo
 \else \expandafter \@secondoftwo
 \fi
}%
\providecommand \@ifx [1]{%
 \ifx #1\expandafter \@firstoftwo
 \else \expandafter \@secondoftwo
 \fi
}%
\providecommand \natexlab [1]{#1}%
\providecommand \enquote  [1]{``#1''}%
\providecommand \bibnamefont  [1]{#1}%
\providecommand \bibfnamefont [1]{#1}%
\providecommand \citenamefont [1]{#1}%
\providecommand \href@noop [0]{\@secondoftwo}%
\providecommand \href [0]{\begingroup \@sanitize@url \@href}%
\providecommand \@href[1]{\@@startlink{#1}\@@href}%
\providecommand \@@href[1]{\endgroup#1\@@endlink}%
\providecommand \@sanitize@url [0]{\catcode `\\12\catcode `\$12\catcode
  `\&12\catcode `\#12\catcode `\^12\catcode `\_12\catcode `\%12\relax}%
\providecommand \@@startlink[1]{}%
\providecommand \@@endlink[0]{}%
\providecommand \url  [0]{\begingroup\@sanitize@url \@url }%
\providecommand \@url [1]{\endgroup\@href {#1}{\urlprefix }}%
\providecommand \urlprefix  [0]{URL }%
\providecommand \Eprint [0]{\href }%
\providecommand \doibase [0]{https://doi.org/}%
\providecommand \selectlanguage [0]{\@gobble}%
\providecommand \bibinfo  [0]{\@secondoftwo}%
\providecommand \bibfield  [0]{\@secondoftwo}%
\providecommand \translation [1]{[#1]}%
\providecommand \BibitemOpen [0]{}%
\providecommand \bibitemStop [0]{}%
\providecommand \bibitemNoStop [0]{.\EOS\space}%
\providecommand \EOS [0]{\spacefactor3000\relax}%
\providecommand \BibitemShut  [1]{\csname bibitem#1\endcsname}%
\let\auto@bib@innerbib\@empty
\bibitem [{\citenamefont {Ryu}\ \emph {et~al.}(2010)\citenamefont {Ryu},
  \citenamefont {Schnyder}, \citenamefont {Furusaki},\ and\ \citenamefont
  {Ludwig}}]{Ryu2010_periodictable}%
  \BibitemOpen
  \bibfield  {author} {\bibinfo {author} {\bibfnamefont {S.}~\bibnamefont
  {Ryu}}, \bibinfo {author} {\bibfnamefont {A.~P.}\ \bibnamefont {Schnyder}},
  \bibinfo {author} {\bibfnamefont {A.}~\bibnamefont {Furusaki}},\ and\
  \bibinfo {author} {\bibfnamefont {A.~W.~W.}\ \bibnamefont {Ludwig}},\
  }\bibfield  {title} {\bibinfo {title} {Topological insulators and
  superconductors: tenfold way and dimensional hierarchy},\ }\href
  {https://doi.org/10.1088/1367-2630/12/6/065010} {\bibfield  {journal}
  {\bibinfo  {journal} {New J. Phys.}\ }\textbf {\bibinfo {volume} {12}},\
  \bibinfo {pages} {065010} (\bibinfo {year} {2010})}\BibitemShut {NoStop}%
\bibitem [{\citenamefont {Thouless}(1983)}]{Thouless1983_prb}%
  \BibitemOpen
  \bibfield  {author} {\bibinfo {author} {\bibfnamefont {D.~J.}\ \bibnamefont
  {Thouless}},\ }\bibfield  {title} {\bibinfo {title} {Quantization of particle
  transport},\ }\href {https://doi.org/10.1103/PhysRevB.27.6083} {\bibfield
  {journal} {\bibinfo  {journal} {Phys. Rev. B}\ }\textbf {\bibinfo {volume}
  {27}},\ \bibinfo {pages} {6083} (\bibinfo {year} {1983})}\BibitemShut
  {NoStop}%
\bibitem [{\citenamefont {den Nijs}\ and\ \citenamefont
  {Rommelse}(1989)}]{Nijs1989}%
  \BibitemOpen
  \bibfield  {author} {\bibinfo {author} {\bibfnamefont {M.}~\bibnamefont {den
  Nijs}}\ and\ \bibinfo {author} {\bibfnamefont {K.}~\bibnamefont {Rommelse}},\
  }\bibfield  {title} {\bibinfo {title} {Preroughening transitions in crystal
  surfaces and valence-bond phases in quantum spin chains},\ }\href
  {https://doi.org/10.1103/PhysRevB.40.4709} {\bibfield  {journal} {\bibinfo
  {journal} {Phys. Rev. B}\ }\textbf {\bibinfo {volume} {40}},\ \bibinfo
  {pages} {4709} (\bibinfo {year} {1989})}\BibitemShut {NoStop}%
\bibitem [{\citenamefont {Citro}\ and\ \citenamefont
  {Aidelsburger}(2023)}]{AidelsburgerThouless2023}%
  \BibitemOpen
  \bibfield  {author} {\bibinfo {author} {\bibfnamefont {R.}~\bibnamefont
  {Citro}}\ and\ \bibinfo {author} {\bibfnamefont {M.}~\bibnamefont
  {Aidelsburger}},\ }\bibfield  {title} {\bibinfo {title} {{T}houless pumping
  and topology},\ }\href {https://doi.org/10.1038/s42254-022-00545-0}
  {\bibfield  {journal} {\bibinfo  {journal} {Nat. Rev. Phys.}\ }\textbf
  {\bibinfo {volume} {5}},\ \bibinfo {pages} {87} (\bibinfo {year}
  {2023})}\BibitemShut {NoStop}%
\bibitem [{\citenamefont {Klitzing}\ \emph {et~al.}(1980)\citenamefont
  {Klitzing}, \citenamefont {Dorda},\ and\ \citenamefont
  {Pepper}}]{Klitzing1980}%
  \BibitemOpen
  \bibfield  {author} {\bibinfo {author} {\bibfnamefont {K.~v.}\ \bibnamefont
  {Klitzing}}, \bibinfo {author} {\bibfnamefont {G.}~\bibnamefont {Dorda}},\
  and\ \bibinfo {author} {\bibfnamefont {M.}~\bibnamefont {Pepper}},\
  }\bibfield  {title} {\bibinfo {title} {{New Method for High-Accuracy
  Determination of the Fine-Structure Constant Based on Quantized Hall
  Resistance}},\ }\href {https://doi.org/10.1103/PhysRevLett.45.494} {\bibfield
   {journal} {\bibinfo  {journal} {Phys. Rev. Lett.}\ }\textbf {\bibinfo
  {volume} {45}},\ \bibinfo {pages} {494} (\bibinfo {year} {1980})}\BibitemShut
  {NoStop}%
\bibitem [{\citenamefont {Thouless}\ \emph {et~al.}(1982)\citenamefont
  {Thouless}, \citenamefont {Kohmoto}, \citenamefont {Nightingale},\ and\
  \citenamefont {den Nijs}}]{TKNN1982}%
  \BibitemOpen
  \bibfield  {author} {\bibinfo {author} {\bibfnamefont {D.~J.}\ \bibnamefont
  {Thouless}}, \bibinfo {author} {\bibfnamefont {M.}~\bibnamefont {Kohmoto}},
  \bibinfo {author} {\bibfnamefont {M.~P.}\ \bibnamefont {Nightingale}},\ and\
  \bibinfo {author} {\bibfnamefont {M.}~\bibnamefont {den Nijs}},\ }\bibfield
  {title} {\bibinfo {title} {{Quantized Hall Conductance in a Two-Dimensional
  Periodic Potential}},\ }\href {https://doi.org/10.1103/PhysRevLett.49.405}
  {\bibfield  {journal} {\bibinfo  {journal} {Phys. Rev. Lett.}\ }\textbf
  {\bibinfo {volume} {49}},\ \bibinfo {pages} {405} (\bibinfo {year}
  {1982})}\BibitemShut {NoStop}%
\bibitem [{\citenamefont {Ozawa}\ \emph {et~al.}(2019)\citenamefont {Ozawa},
  \citenamefont {Price}, \citenamefont {Amo}, \citenamefont {Goldman},
  \citenamefont {Hafezi}, \citenamefont {Lu}, \citenamefont {Rechtsman},
  \citenamefont {Schuster}, \citenamefont {Simon}, \citenamefont {Zilberberg},\
  and\ \citenamefont {Carusotto}}]{TopologicalPhotonicsreview_2019}%
  \BibitemOpen
  \bibfield  {author} {\bibinfo {author} {\bibfnamefont {T.}~\bibnamefont
  {Ozawa}}, \bibinfo {author} {\bibfnamefont {H.~M.}\ \bibnamefont {Price}},
  \bibinfo {author} {\bibfnamefont {A.}~\bibnamefont {Amo}}, \bibinfo {author}
  {\bibfnamefont {N.}~\bibnamefont {Goldman}}, \bibinfo {author} {\bibfnamefont
  {M.}~\bibnamefont {Hafezi}}, \bibinfo {author} {\bibfnamefont
  {L.}~\bibnamefont {Lu}}, \bibinfo {author} {\bibfnamefont {M.~C.}\
  \bibnamefont {Rechtsman}}, \bibinfo {author} {\bibfnamefont {D.}~\bibnamefont
  {Schuster}}, \bibinfo {author} {\bibfnamefont {J.}~\bibnamefont {Simon}},
  \bibinfo {author} {\bibfnamefont {O.}~\bibnamefont {Zilberberg}},\ and\
  \bibinfo {author} {\bibfnamefont {I.}~\bibnamefont {Carusotto}},\ }\bibfield
  {title} {\bibinfo {title} {Topological photonics},\ }\href
  {https://doi.org/10.1103/RevModPhys.91.015006} {\bibfield  {journal}
  {\bibinfo  {journal} {Rev. Mod. Phys.}\ }\textbf {\bibinfo {volume} {91}},\
  \bibinfo {pages} {015006} (\bibinfo {year} {2019})}\BibitemShut {NoStop}%
\bibitem [{\citenamefont {Cerjan}\ \emph {et~al.}(2020)\citenamefont {Cerjan},
  \citenamefont {Wang}, \citenamefont {Huang}, \citenamefont {Chen},\ and\
  \citenamefont {Rechtsman}}]{Cerjan2020Thouless}%
  \BibitemOpen
  \bibfield  {author} {\bibinfo {author} {\bibfnamefont {A.}~\bibnamefont
  {Cerjan}}, \bibinfo {author} {\bibfnamefont {M.}~\bibnamefont {Wang}},
  \bibinfo {author} {\bibfnamefont {S.}~\bibnamefont {Huang}}, \bibinfo
  {author} {\bibfnamefont {K.~P.}\ \bibnamefont {Chen}},\ and\ \bibinfo
  {author} {\bibfnamefont {M.~C.}\ \bibnamefont {Rechtsman}},\ }\bibfield
  {title} {\bibinfo {title} {{T}houless pumping in disordered photonic
  systems},\ }\href {https://doi.org/10.1038/s41377-020-00408-2} {\bibfield
  {journal} {\bibinfo  {journal} {Light: Science \& Applications}\ }\textbf
  {\bibinfo {volume} {9}},\ \bibinfo {pages} {178} (\bibinfo {year}
  {2020})}\BibitemShut {NoStop}%
\bibitem [{\citenamefont {J{\"u}rgensen}\ \emph {et~al.}(2021)\citenamefont
  {J{\"u}rgensen}, \citenamefont {Mukherjee},\ and\ \citenamefont
  {Rechtsman}}]{Jurguensen2021Thouless}%
  \BibitemOpen
  \bibfield  {author} {\bibinfo {author} {\bibfnamefont {M.}~\bibnamefont
  {J{\"u}rgensen}}, \bibinfo {author} {\bibfnamefont {S.}~\bibnamefont
  {Mukherjee}},\ and\ \bibinfo {author} {\bibfnamefont {M.~C.}\ \bibnamefont
  {Rechtsman}},\ }\bibfield  {title} {\bibinfo {title} {Quantized nonlinear
  {T}houless pumping},\ }\href {https://doi.org/10.1038/s41586-021-03688-9}
  {\bibfield  {journal} {\bibinfo  {journal} {Nature}\ }\textbf {\bibinfo
  {volume} {596}},\ \bibinfo {pages} {63} (\bibinfo {year} {2021})}\BibitemShut
  {NoStop}%
\bibitem [{\citenamefont {Kraus}\ \emph {et~al.}(2012)\citenamefont {Kraus},
  \citenamefont {Lahini}, \citenamefont {Ringel}, \citenamefont {Verbin},\ and\
  \citenamefont {Zilberberg}}]{Kraus2012_pump}%
  \BibitemOpen
  \bibfield  {author} {\bibinfo {author} {\bibfnamefont {Y.~E.}\ \bibnamefont
  {Kraus}}, \bibinfo {author} {\bibfnamefont {Y.}~\bibnamefont {Lahini}},
  \bibinfo {author} {\bibfnamefont {Z.}~\bibnamefont {Ringel}}, \bibinfo
  {author} {\bibfnamefont {M.}~\bibnamefont {Verbin}},\ and\ \bibinfo {author}
  {\bibfnamefont {O.}~\bibnamefont {Zilberberg}},\ }\bibfield  {title}
  {\bibinfo {title} {Topological states and adiabatic pumping in
  quasicrystals},\ }\href {https://doi.org/10.1103/PhysRevLett.109.106402}
  {\bibfield  {journal} {\bibinfo  {journal} {Phys. Rev. Lett.}\ }\textbf
  {\bibinfo {volume} {109}},\ \bibinfo {pages} {106402} (\bibinfo {year}
  {2012})}\BibitemShut {NoStop}%
\bibitem [{\citenamefont {Lewenstein}\ \emph {et~al.}(2012)\citenamefont
  {Lewenstein}, \citenamefont {Sanpera},\ and\ \citenamefont
  {Ahufinger}}]{MaciejBook}%
  \BibitemOpen
  \bibfield  {author} {\bibinfo {author} {\bibfnamefont {M.}~\bibnamefont
  {Lewenstein}}, \bibinfo {author} {\bibfnamefont {A.}~\bibnamefont
  {Sanpera}},\ and\ \bibinfo {author} {\bibfnamefont {V.}~\bibnamefont
  {Ahufinger}},\ }\href
  {http://www.oxfordscholarship.com/view/10.1093/acprof:oso/9780199573127.001.0001/acprof-9780199573127}
  {\emph {\bibinfo {title} {Ultracold Atoms in Optical Lattices: Simulating
  Quantum many-body systems}}},\ Vol.~\bibinfo {volume} {54}\ (\bibinfo
  {publisher} {Oxford University Press, Oxford},\ \bibinfo {year}
  {2012})\BibitemShut {NoStop}%
\bibitem [{\citenamefont {Bloch}\ \emph {et~al.}(2008)\citenamefont {Bloch},
  \citenamefont {Dalibard},\ and\ \citenamefont {Zwerger}}]{Bloch_2008}%
  \BibitemOpen
  \bibfield  {author} {\bibinfo {author} {\bibfnamefont {I.}~\bibnamefont
  {Bloch}}, \bibinfo {author} {\bibfnamefont {J.}~\bibnamefont {Dalibard}},\
  and\ \bibinfo {author} {\bibfnamefont {W.}~\bibnamefont {Zwerger}},\
  }\bibfield  {title} {\bibinfo {title} {Many-body physics with ultracold
  gases},\ }\href {https://doi.org/10.1103/RevModPhys.80.885} {\bibfield
  {journal} {\bibinfo  {journal} {Rev. Mod. Phys.}\ }\textbf {\bibinfo {volume}
  {80}},\ \bibinfo {pages} {885} (\bibinfo {year} {2008})}\BibitemShut
  {NoStop}%
\bibitem [{\citenamefont {Cooper}\ \emph {et~al.}(2019)\citenamefont {Cooper},
  \citenamefont {Dalibard},\ and\ \citenamefont {Spielman}}]{Cooper_2019}%
  \BibitemOpen
  \bibfield  {author} {\bibinfo {author} {\bibfnamefont {N.~R.}\ \bibnamefont
  {Cooper}}, \bibinfo {author} {\bibfnamefont {J.}~\bibnamefont {Dalibard}},\
  and\ \bibinfo {author} {\bibfnamefont {I.~B.}\ \bibnamefont {Spielman}},\
  }\bibfield  {title} {\bibinfo {title} {Topological bands for ultracold
  atoms},\ }\href {https://doi.org/10.1103/RevModPhys.91.015005} {\bibfield
  {journal} {\bibinfo  {journal} {Rev. Mod. Phys.}\ }\textbf {\bibinfo {volume}
  {91}},\ \bibinfo {pages} {015005} (\bibinfo {year} {2019})}\BibitemShut
  {NoStop}%
\bibitem [{\citenamefont {Nakajima}\ \emph {et~al.}(2016)\citenamefont
  {Nakajima}, \citenamefont {Tomita}, \citenamefont {Taie}, \citenamefont
  {Ichinose}, \citenamefont {Ozawa}, \citenamefont {Wang}, \citenamefont
  {Troyer},\ and\ \citenamefont {Takahashi}}]{Nakajima2016}%
  \BibitemOpen
  \bibfield  {author} {\bibinfo {author} {\bibfnamefont {S.}~\bibnamefont
  {Nakajima}}, \bibinfo {author} {\bibfnamefont {T.}~\bibnamefont {Tomita}},
  \bibinfo {author} {\bibfnamefont {S.}~\bibnamefont {Taie}}, \bibinfo {author}
  {\bibfnamefont {T.}~\bibnamefont {Ichinose}}, \bibinfo {author}
  {\bibfnamefont {H.}~\bibnamefont {Ozawa}}, \bibinfo {author} {\bibfnamefont
  {L.}~\bibnamefont {Wang}}, \bibinfo {author} {\bibfnamefont {M.}~\bibnamefont
  {Troyer}},\ and\ \bibinfo {author} {\bibfnamefont {Y.}~\bibnamefont
  {Takahashi}},\ }\bibfield  {title} {\bibinfo {title} {Topological {T}houless
  pumping of ultracold fermions},\ }\href {https://doi.org/10.1038/nphys3622}
  {\bibfield  {journal} {\bibinfo  {journal} {Nat. Phys.}\ }\textbf {\bibinfo
  {volume} {12}},\ \bibinfo {pages} {296} (\bibinfo {year} {2016})}\BibitemShut
  {NoStop}%
\bibitem [{\citenamefont {Minguzzi}\ \emph {et~al.}(2022)\citenamefont
  {Minguzzi}, \citenamefont {Zhu}, \citenamefont {Sandholzer}, \citenamefont
  {Walter}, \citenamefont {Viebahn},\ and\ \citenamefont
  {Esslinger}}]{Minguzzi2022}%
  \BibitemOpen
  \bibfield  {author} {\bibinfo {author} {\bibfnamefont {J.}~\bibnamefont
  {Minguzzi}}, \bibinfo {author} {\bibfnamefont {Z.}~\bibnamefont {Zhu}},
  \bibinfo {author} {\bibfnamefont {K.}~\bibnamefont {Sandholzer}}, \bibinfo
  {author} {\bibfnamefont {A.-S.}\ \bibnamefont {Walter}}, \bibinfo {author}
  {\bibfnamefont {K.}~\bibnamefont {Viebahn}},\ and\ \bibinfo {author}
  {\bibfnamefont {T.}~\bibnamefont {Esslinger}},\ }\bibfield  {title} {\bibinfo
  {title} {{Topological Pumping in a Floquet-Bloch Band}},\ }\href
  {https://doi.org/10.1103/PhysRevLett.129.053201} {\bibfield  {journal}
  {\bibinfo  {journal} {Phys. Rev. Lett.}\ }\textbf {\bibinfo {volume} {129}},\
  \bibinfo {pages} {053201} (\bibinfo {year} {2022})}\BibitemShut {NoStop}%
\bibitem [{\citenamefont {Lohse}\ \emph {et~al.}(2016)\citenamefont {Lohse},
  \citenamefont {Schweizer}, \citenamefont {Zilberberg}, \citenamefont
  {Aidelsburger},\ and\ \citenamefont {Bloch}}]{Lohse2016}%
  \BibitemOpen
  \bibfield  {author} {\bibinfo {author} {\bibfnamefont {M.}~\bibnamefont
  {Lohse}}, \bibinfo {author} {\bibfnamefont {C.}~\bibnamefont {Schweizer}},
  \bibinfo {author} {\bibfnamefont {O.}~\bibnamefont {Zilberberg}}, \bibinfo
  {author} {\bibfnamefont {M.}~\bibnamefont {Aidelsburger}},\ and\ \bibinfo
  {author} {\bibfnamefont {I.}~\bibnamefont {Bloch}},\ }\bibfield  {title}
  {\bibinfo {title} {A {T}houless quantum pump with ultracold bosonic atoms in
  an optical superlattice},\ }\href {https://doi.org/0.1038/nphys3584}
  {\bibfield  {journal} {\bibinfo  {journal} {Nat. Phys.}\ }\textbf {\bibinfo
  {volume} {12}},\ \bibinfo {pages} {350} (\bibinfo {year} {2016})}\BibitemShut
  {NoStop}%
\bibitem [{\citenamefont {Rice}\ and\ \citenamefont
  {Mele}(1982)}]{RiceMele1982}%
  \BibitemOpen
  \bibfield  {author} {\bibinfo {author} {\bibfnamefont {M.~J.}\ \bibnamefont
  {Rice}}\ and\ \bibinfo {author} {\bibfnamefont {E.~J.}\ \bibnamefont
  {Mele}},\ }\bibfield  {title} {\bibinfo {title} {Elementary excitations of a
  linearly conjugated diatomic polymer},\ }\href
  {https://doi.org/10.1103/PhysRevLett.49.1455} {\bibfield  {journal} {\bibinfo
   {journal} {Phys. Rev. Lett.}\ }\textbf {\bibinfo {volume} {49}},\ \bibinfo
  {pages} {1455} (\bibinfo {year} {1982})}\BibitemShut {NoStop}%
\bibitem [{\citenamefont {Rachel}\ and\ \citenamefont
  {Le~Hur}(2010)}]{rachel_2010}%
  \BibitemOpen
  \bibfield  {author} {\bibinfo {author} {\bibfnamefont {S.}~\bibnamefont
  {Rachel}}\ and\ \bibinfo {author} {\bibfnamefont {K.}~\bibnamefont
  {Le~Hur}},\ }\bibfield  {title} {\bibinfo {title} {{Topological insulators
  and Mott physics from the {H}ubbard interaction}},\ }\href
  {https://doi.org/10.1103/PhysRevB.82.075106} {\bibfield  {journal} {\bibinfo
  {journal} {Phys. Rev. B}\ }\textbf {\bibinfo {volume} {82}},\ \bibinfo
  {pages} {075106} (\bibinfo {year} {2010})}\BibitemShut {NoStop}%
\bibitem [{\citenamefont {Berg}\ \emph {et~al.}(2011)\citenamefont {Berg},
  \citenamefont {Levin},\ and\ \citenamefont {Altman}}]{Berg2011}%
  \BibitemOpen
  \bibfield  {author} {\bibinfo {author} {\bibfnamefont {E.}~\bibnamefont
  {Berg}}, \bibinfo {author} {\bibfnamefont {M.}~\bibnamefont {Levin}},\ and\
  \bibinfo {author} {\bibfnamefont {E.}~\bibnamefont {Altman}},\ }\bibfield
  {title} {\bibinfo {title} {Quantized pumping and topology of the phase
  diagram for a system of interacting bosons},\ }\href
  {https://doi.org/10.1103/PhysRevLett.106.110405} {\bibfield  {journal}
  {\bibinfo  {journal} {Phys. Rev. Lett.}\ }\textbf {\bibinfo {volume} {106}},\
  \bibinfo {pages} {110405} (\bibinfo {year} {2011})}\BibitemShut {NoStop}%
\bibitem [{\citenamefont {Greschner}\ \emph {et~al.}(2020)\citenamefont
  {Greschner}, \citenamefont {Mondal},\ and\ \citenamefont
  {Mishra}}]{Greschner2020}%
  \BibitemOpen
  \bibfield  {author} {\bibinfo {author} {\bibfnamefont {S.}~\bibnamefont
  {Greschner}}, \bibinfo {author} {\bibfnamefont {S.}~\bibnamefont {Mondal}},\
  and\ \bibinfo {author} {\bibfnamefont {T.}~\bibnamefont {Mishra}},\
  }\bibfield  {title} {\bibinfo {title} {Topological charge pumping of bound
  bosonic pairs},\ }\href {https://doi.org/10.1103/PhysRevA.101.053630}
  {\bibfield  {journal} {\bibinfo  {journal} {Phys. Rev. A}\ }\textbf {\bibinfo
  {volume} {101}},\ \bibinfo {pages} {053630} (\bibinfo {year}
  {2020})}\BibitemShut {NoStop}%
\bibitem [{\citenamefont {Hayward}\ \emph {et~al.}(2018)\citenamefont
  {Hayward}, \citenamefont {Schweizer}, \citenamefont {Lohse}, \citenamefont
  {Aidelsburger},\ and\ \citenamefont {Heidrich-Meisner}}]{hayward2018}%
  \BibitemOpen
  \bibfield  {author} {\bibinfo {author} {\bibfnamefont {A.}~\bibnamefont
  {Hayward}}, \bibinfo {author} {\bibfnamefont {C.}~\bibnamefont {Schweizer}},
  \bibinfo {author} {\bibfnamefont {M.}~\bibnamefont {Lohse}}, \bibinfo
  {author} {\bibfnamefont {M.}~\bibnamefont {Aidelsburger}},\ and\ \bibinfo
  {author} {\bibfnamefont {F.}~\bibnamefont {Heidrich-Meisner}},\ }\bibfield
  {title} {\bibinfo {title} {Topological charge pumping in the interacting
  bosonic {Rice-Mele} model},\ }\href
  {https://doi.org/10.1103/PhysRevB.98.245148} {\bibfield  {journal} {\bibinfo
  {journal} {Phys. Rev. B}\ }\textbf {\bibinfo {volume} {98}},\ \bibinfo
  {pages} {245148} (\bibinfo {year} {2018})}\BibitemShut {NoStop}%
\bibitem [{\citenamefont {Mondal}\ \emph {et~al.}(2021)\citenamefont {Mondal},
  \citenamefont {Greschner}, \citenamefont {Santos},\ and\ \citenamefont
  {Mishra}}]{mondal2021_inheritance}%
  \BibitemOpen
  \bibfield  {author} {\bibinfo {author} {\bibfnamefont {S.}~\bibnamefont
  {Mondal}}, \bibinfo {author} {\bibfnamefont {S.}~\bibnamefont {Greschner}},
  \bibinfo {author} {\bibfnamefont {L.}~\bibnamefont {Santos}},\ and\ \bibinfo
  {author} {\bibfnamefont {T.}~\bibnamefont {Mishra}},\ }\bibfield  {title}
  {\bibinfo {title} {{Topological inheritance in two-component {H}ubbard models
  with single-component Su-Schrieffer-Heeger dimerization}},\ }\href
  {https://doi.org/10.1103/PhysRevA.104.013315} {\bibfield  {journal} {\bibinfo
   {journal} {Phys. Rev. A}\ }\textbf {\bibinfo {volume} {104}},\ \bibinfo
  {pages} {013315} (\bibinfo {year} {2021})}\BibitemShut {NoStop}%
\bibitem [{\citenamefont {Lin}\ \emph {et~al.}(2020)\citenamefont {Lin},
  \citenamefont {Ke},\ and\ \citenamefont {Lee}}]{Lin2020_dimerized}%
  \BibitemOpen
  \bibfield  {author} {\bibinfo {author} {\bibfnamefont {L.}~\bibnamefont
  {Lin}}, \bibinfo {author} {\bibfnamefont {Y.}~\bibnamefont {Ke}},\ and\
  \bibinfo {author} {\bibfnamefont {C.}~\bibnamefont {Lee}},\ }\bibfield
  {title} {\bibinfo {title} {Interaction-induced topological bound states and
  {T}houless pumping in a one-dimensional optical lattice},\ }\href
  {https://doi.org/10.1103/PhysRevA.101.023620} {\bibfield  {journal} {\bibinfo
   {journal} {Phys. Rev. A}\ }\textbf {\bibinfo {volume} {101}},\ \bibinfo
  {pages} {023620} (\bibinfo {year} {2020})}\BibitemShut {NoStop}%
\bibitem [{\citenamefont {Mondal}\ \emph {et~al.}(2022)\citenamefont {Mondal},
  \citenamefont {Padhan},\ and\ \citenamefont {Mishra}}]{Mondal2022}%
  \BibitemOpen
  \bibfield  {author} {\bibinfo {author} {\bibfnamefont {S.}~\bibnamefont
  {Mondal}}, \bibinfo {author} {\bibfnamefont {A.}~\bibnamefont {Padhan}},\
  and\ \bibinfo {author} {\bibfnamefont {T.}~\bibnamefont {Mishra}},\
  }\bibfield  {title} {\bibinfo {title} {Realizing a symmetry protected
  topological phase through dimerized interactions},\ }\href
  {https://doi.org/10.1103/PhysRevB.106.L201106} {\bibfield  {journal}
  {\bibinfo  {journal} {Phys. Rev. B}\ }\textbf {\bibinfo {volume} {106}},\
  \bibinfo {pages} {L201106} (\bibinfo {year} {2022})}\BibitemShut {NoStop}%
\bibitem [{\citenamefont {Kuno}\ and\ \citenamefont
  {Hatsugai}(2020)}]{Kuno2020}%
  \BibitemOpen
  \bibfield  {author} {\bibinfo {author} {\bibfnamefont {Y.}~\bibnamefont
  {Kuno}}\ and\ \bibinfo {author} {\bibfnamefont {Y.}~\bibnamefont
  {Hatsugai}},\ }\bibfield  {title} {\bibinfo {title} {Interaction-induced
  topological charge pump},\ }\href
  {https://doi.org/10.1103/PhysRevResearch.2.042024} {\bibfield  {journal}
  {\bibinfo  {journal} {Phys. Rev. Res.}\ }\textbf {\bibinfo {volume} {2}},\
  \bibinfo {pages} {042024} (\bibinfo {year} {2020})}\BibitemShut {NoStop}%
\bibitem [{\citenamefont {Padhan}\ \emph {et~al.}(2024)\citenamefont {Padhan},
  \citenamefont {Mondal}, \citenamefont {Vishveshwara},\ and\ \citenamefont
  {Mishra}}]{padhan2023interacting}%
  \BibitemOpen
  \bibfield  {author} {\bibinfo {author} {\bibfnamefont {A.}~\bibnamefont
  {Padhan}}, \bibinfo {author} {\bibfnamefont {S.}~\bibnamefont {Mondal}},
  \bibinfo {author} {\bibfnamefont {S.}~\bibnamefont {Vishveshwara}},\ and\
  \bibinfo {author} {\bibfnamefont {T.}~\bibnamefont {Mishra}},\ }\bibfield
  {title} {\bibinfo {title} {Interacting bosons on a su-schrieffer-heeger
  ladder: Topological phases and thouless pumping},\ }\href
  {https://doi.org/10.1103/PhysRevB.109.085120} {\bibfield  {journal} {\bibinfo
   {journal} {Phys. Rev. B}\ }\textbf {\bibinfo {volume} {109}},\ \bibinfo
  {pages} {085120} (\bibinfo {year} {2024})}\BibitemShut {NoStop}%
\bibitem [{\citenamefont {Athanasiou}\ \emph {et~al.}(2024)\citenamefont
  {Athanasiou}, \citenamefont {Nielsen}, \citenamefont {Wauters},\ and\
  \citenamefont {Burrello}}]{athanasiou2023thouless}%
  \BibitemOpen
  \bibfield  {author} {\bibinfo {author} {\bibfnamefont {S.}~\bibnamefont
  {Athanasiou}}, \bibinfo {author} {\bibfnamefont {I.~E.}\ \bibnamefont
  {Nielsen}}, \bibinfo {author} {\bibfnamefont {M.~M.}\ \bibnamefont
  {Wauters}},\ and\ \bibinfo {author} {\bibfnamefont {M.}~\bibnamefont
  {Burrello}},\ }\bibfield  {title} {\bibinfo {title} {{Thouless pumping in
  Josephson junction arrays}},\ }\href
  {https://doi.org/10.21468/SciPostPhys.16.3.083} {\bibfield  {journal}
  {\bibinfo  {journal} {SciPost Phys.}\ }\textbf {\bibinfo {volume} {16}},\
  \bibinfo {pages} {083} (\bibinfo {year} {2024})}\BibitemShut {NoStop}%
\bibitem [{\citenamefont {Nakagawa}\ \emph {et~al.}(2018)\citenamefont
  {Nakagawa}, \citenamefont {Yoshida}, \citenamefont {Peters},\ and\
  \citenamefont {Kawakami}}]{Nakagawa2018}%
  \BibitemOpen
  \bibfield  {author} {\bibinfo {author} {\bibfnamefont {M.}~\bibnamefont
  {Nakagawa}}, \bibinfo {author} {\bibfnamefont {T.}~\bibnamefont {Yoshida}},
  \bibinfo {author} {\bibfnamefont {R.}~\bibnamefont {Peters}},\ and\ \bibinfo
  {author} {\bibfnamefont {N.}~\bibnamefont {Kawakami}},\ }\bibfield  {title}
  {\bibinfo {title} {Breakdown of topological {T}houless pumping in the
  strongly interacting regime},\ }\href
  {https://doi.org/10.1103/PhysRevB.98.115147} {\bibfield  {journal} {\bibinfo
  {journal} {Phys. Rev. B}\ }\textbf {\bibinfo {volume} {98}},\ \bibinfo
  {pages} {115147} (\bibinfo {year} {2018})}\BibitemShut {NoStop}%
\bibitem [{\citenamefont {Bertok}\ \emph {et~al.}(2022)\citenamefont {Bertok},
  \citenamefont {Heidrich-Meisner},\ and\ \citenamefont {Aligia}}]{Bertok2022}%
  \BibitemOpen
  \bibfield  {author} {\bibinfo {author} {\bibfnamefont {E.}~\bibnamefont
  {Bertok}}, \bibinfo {author} {\bibfnamefont {F.}~\bibnamefont
  {Heidrich-Meisner}},\ and\ \bibinfo {author} {\bibfnamefont {A.~A.}\
  \bibnamefont {Aligia}},\ }\bibfield  {title} {\bibinfo {title} {Splitting of
  topological charge pumping in an interacting two-component fermionic
  {R}ice-{M}ele {H}ubbard model},\ }\href
  {https://doi.org/10.1103/PhysRevB.106.045141} {\bibfield  {journal} {\bibinfo
   {journal} {Phys. Rev. B}\ }\textbf {\bibinfo {volume} {106}},\ \bibinfo
  {pages} {045141} (\bibinfo {year} {2022})}\BibitemShut {NoStop}%
\bibitem [{\citenamefont {Walter}\ \emph {et~al.}(2023)\citenamefont {Walter},
  \citenamefont {Zhu}, \citenamefont {G{\"a}chter}, \citenamefont {Minguzzi},
  \citenamefont {Roschinski}, \citenamefont {Sandholzer}, \citenamefont
  {Viebahn},\ and\ \citenamefont {Esslinger}}]{Walter2022}%
  \BibitemOpen
  \bibfield  {author} {\bibinfo {author} {\bibfnamefont {A.-S.}\ \bibnamefont
  {Walter}}, \bibinfo {author} {\bibfnamefont {Z.}~\bibnamefont {Zhu}},
  \bibinfo {author} {\bibfnamefont {M.}~\bibnamefont {G{\"a}chter}}, \bibinfo
  {author} {\bibfnamefont {J.}~\bibnamefont {Minguzzi}}, \bibinfo {author}
  {\bibfnamefont {S.}~\bibnamefont {Roschinski}}, \bibinfo {author}
  {\bibfnamefont {K.}~\bibnamefont {Sandholzer}}, \bibinfo {author}
  {\bibfnamefont {K.}~\bibnamefont {Viebahn}},\ and\ \bibinfo {author}
  {\bibfnamefont {T.}~\bibnamefont {Esslinger}},\ }\bibfield  {title} {\bibinfo
  {title} {Quantization and its breakdown in a hubbard--thouless pump},\ }\href
  {https://doi.org/10.1038/s41567-023-02145-w} {\bibfield  {journal} {\bibinfo
  {journal} {Nat. Phys.}\ }\textbf {\bibinfo {volume} {19}},\ \bibinfo {pages}
  {1471} (\bibinfo {year} {2023})}\BibitemShut {NoStop}%
\bibitem [{\citenamefont {Arg{\"{u}}ello-Luengo}\ \emph
  {et~al.}(2024)\citenamefont {Arg{\"{u}}ello-Luengo}, \citenamefont {Mark},
  \citenamefont {Ferlaino}, \citenamefont {Lewenstein}, \citenamefont
  {Barbiero},\ and\ \citenamefont
  {Juli{\`{a}}-Farr{\'{e}}}}]{arguello2023stabilization}%
  \BibitemOpen
  \bibfield  {author} {\bibinfo {author} {\bibfnamefont {J.}~\bibnamefont
  {Arg{\"{u}}ello-Luengo}}, \bibinfo {author} {\bibfnamefont {M.~J.}\
  \bibnamefont {Mark}}, \bibinfo {author} {\bibfnamefont {F.}~\bibnamefont
  {Ferlaino}}, \bibinfo {author} {\bibfnamefont {M.}~\bibnamefont
  {Lewenstein}}, \bibinfo {author} {\bibfnamefont {L.}~\bibnamefont
  {Barbiero}},\ and\ \bibinfo {author} {\bibfnamefont {S.}~\bibnamefont
  {Juli{\`{a}}-Farr{\'{e}}}},\ }\bibfield  {title} {\bibinfo {title}
  {Stabilization of {H}ubbard-{T}houless pumps through nonlocal fermionic
  repulsion},\ }\href {https://doi.org/10.22331/q-2024-03-14-1285} {\bibfield
  {journal} {\bibinfo  {journal} {{Quantum}}\ }\textbf {\bibinfo {volume}
  {8}},\ \bibinfo {pages} {1285} (\bibinfo {year} {2024})}\BibitemShut
  {NoStop}%
\bibitem [{\citenamefont {Segura}\ \emph {et~al.}(2023)\citenamefont {Segura},
  \citenamefont {Hallberg},\ and\ \citenamefont {Aligia}}]{segura2023charge}%
  \BibitemOpen
  \bibfield  {author} {\bibinfo {author} {\bibfnamefont {O.~A.~M.}\
  \bibnamefont {Segura}}, \bibinfo {author} {\bibfnamefont {K.}~\bibnamefont
  {Hallberg}},\ and\ \bibinfo {author} {\bibfnamefont {A.~A.}\ \bibnamefont
  {Aligia}},\ }\bibfield  {title} {\bibinfo {title} {Charge and spin gaps in
  the ionic hubbard model with density-dependent hopping},\ }\href
  {https://doi.org/10.1103/PhysRevB.108.195135} {\bibfield  {journal} {\bibinfo
   {journal} {Phys. Rev. B}\ }\textbf {\bibinfo {volume} {108}},\ \bibinfo
  {pages} {195135} (\bibinfo {year} {2023})}\BibitemShut {NoStop}%
\bibitem [{\citenamefont {Nakamura}(1999)}]{Nakamura99}%
  \BibitemOpen
  \bibfield  {author} {\bibinfo {author} {\bibfnamefont {M.}~\bibnamefont
  {Nakamura}},\ }\bibfield  {title} {\bibinfo {title} {{Mechanism of CDW-SDW
  Transition in One Dimension}},\ }\href {https://doi.org/10.1143/JPSJ.68.3123}
  {\bibfield  {journal} {\bibinfo  {journal} {J. Phys. Soc. Japan}\ }\textbf
  {\bibinfo {volume} {68}},\ \bibinfo {pages} {3123} (\bibinfo {year}
  {1999})}\BibitemShut {NoStop}%
\bibitem [{\citenamefont {Nakamura}(2000)}]{Nakamura2000}%
  \BibitemOpen
  \bibfield  {author} {\bibinfo {author} {\bibfnamefont {M.}~\bibnamefont
  {Nakamura}},\ }\bibfield  {title} {\bibinfo {title} {Tricritical behavior in
  the extended {H}ubbard chains},\ }\href
  {https://doi.org/10.1103/PhysRevB.61.16377} {\bibfield  {journal} {\bibinfo
  {journal} {Phys. Rev. B}\ }\textbf {\bibinfo {volume} {61}},\ \bibinfo
  {pages} {16377} (\bibinfo {year} {2000})}\BibitemShut {NoStop}%
\bibitem [{\citenamefont {Juli\`a-Farr\'e}\ \emph {et~al.}(2022)\citenamefont
  {Juli\`a-Farr\'e}, \citenamefont {Gonz\'alez-Cuadra}, \citenamefont
  {Patscheider}, \citenamefont {Mark}, \citenamefont {Ferlaino}, \citenamefont
  {Lewenstein}, \citenamefont {Barbiero},\ and\ \citenamefont
  {Dauphin}}]{julia-farre2021revealing}%
  \BibitemOpen
  \bibfield  {author} {\bibinfo {author} {\bibfnamefont {S.}~\bibnamefont
  {Juli\`a-Farr\'e}}, \bibinfo {author} {\bibfnamefont {D.}~\bibnamefont
  {Gonz\'alez-Cuadra}}, \bibinfo {author} {\bibfnamefont {A.}~\bibnamefont
  {Patscheider}}, \bibinfo {author} {\bibfnamefont {M.~J.}\ \bibnamefont
  {Mark}}, \bibinfo {author} {\bibfnamefont {F.}~\bibnamefont {Ferlaino}},
  \bibinfo {author} {\bibfnamefont {M.}~\bibnamefont {Lewenstein}}, \bibinfo
  {author} {\bibfnamefont {L.}~\bibnamefont {Barbiero}},\ and\ \bibinfo
  {author} {\bibfnamefont {A.}~\bibnamefont {Dauphin}},\ }\bibfield  {title}
  {\bibinfo {title} {Revealing the topological nature of the bond order wave in
  a strongly correlated quantum system},\ }\href
  {https://doi.org/10.1103/PhysRevResearch.4.L032005} {\bibfield  {journal}
  {\bibinfo  {journal} {Phys. Rev. Res.}\ }\textbf {\bibinfo {volume} {4}},\
  \bibinfo {pages} {L032005} (\bibinfo {year} {2022})}\BibitemShut {NoStop}%
\bibitem [{\citenamefont {Viebahn}\ \emph {et~al.}(2024)\citenamefont
  {Viebahn}, \citenamefont {Walter}, \citenamefont {Bertok}, \citenamefont
  {Zhu}, \citenamefont {G\"achter}, \citenamefont {Aligia}, \citenamefont
  {Heidrich-Meisner},\ and\ \citenamefont
  {Esslinger}}]{viebahn2023interactioninduced}%
  \BibitemOpen
  \bibfield  {author} {\bibinfo {author} {\bibfnamefont {K.}~\bibnamefont
  {Viebahn}}, \bibinfo {author} {\bibfnamefont {A.-S.}\ \bibnamefont {Walter}},
  \bibinfo {author} {\bibfnamefont {E.}~\bibnamefont {Bertok}}, \bibinfo
  {author} {\bibfnamefont {Z.}~\bibnamefont {Zhu}}, \bibinfo {author}
  {\bibfnamefont {M.}~\bibnamefont {G\"achter}}, \bibinfo {author}
  {\bibfnamefont {A.~A.}\ \bibnamefont {Aligia}}, \bibinfo {author}
  {\bibfnamefont {F.}~\bibnamefont {Heidrich-Meisner}},\ and\ \bibinfo {author}
  {\bibfnamefont {T.}~\bibnamefont {Esslinger}},\ }\bibfield  {title} {\bibinfo
  {title} {Interactions enable thouless pumping in a nonsliding lattice},\
  }\href {https://doi.org/10.1103/PhysRevX.14.021049} {\bibfield  {journal}
  {\bibinfo  {journal} {Phys. Rev. X}\ }\textbf {\bibinfo {volume} {14}},\
  \bibinfo {pages} {021049} (\bibinfo {year} {2024})}\BibitemShut {NoStop}%
\bibitem [{\citenamefont {Elben}\ \emph {et~al.}(2020)\citenamefont {Elben},
  \citenamefont {Yu}, \citenamefont {Zhu}, \citenamefont {Hafezi},
  \citenamefont {Pollmann}, \citenamefont {Zoller},\ and\ \citenamefont
  {Vermersch}}]{elben2020}%
  \BibitemOpen
  \bibfield  {author} {\bibinfo {author} {\bibfnamefont {A.}~\bibnamefont
  {Elben}}, \bibinfo {author} {\bibfnamefont {J.}~\bibnamefont {Yu}}, \bibinfo
  {author} {\bibfnamefont {G.}~\bibnamefont {Zhu}}, \bibinfo {author}
  {\bibfnamefont {M.}~\bibnamefont {Hafezi}}, \bibinfo {author} {\bibfnamefont
  {F.}~\bibnamefont {Pollmann}}, \bibinfo {author} {\bibfnamefont
  {P.}~\bibnamefont {Zoller}},\ and\ \bibinfo {author} {\bibfnamefont
  {B.}~\bibnamefont {Vermersch}},\ }\bibfield  {title} {\bibinfo {title}
  {Many-body topological invariants from randomized measurements in synthetic
  quantum matter},\ }\href {https://doi.org/10.1126/sciadv.aaz3666} {\bibfield
  {journal} {\bibinfo  {journal} {Sci. Adv.}\ }\textbf {\bibinfo {volume}
  {6}},\ \bibinfo {pages} {eaaz3666} (\bibinfo {year} {2020})}\BibitemShut
  {NoStop}%
\bibitem [{\citenamefont {Schollwöck}(2011)}]{schollwock_density-matrix_2011}%
  \BibitemOpen
  \bibfield  {author} {\bibinfo {author} {\bibfnamefont {U.}~\bibnamefont
  {Schollwöck}},\ }\bibfield  {title} {\bibinfo {title} {The density-matrix
  renormalization group in the age of matrix product states},\ }\href
  {https://doi.org/10.1016/j.aop.2010.09.012} {\bibfield  {journal} {\bibinfo
  {journal} {Ann. Phys.}\ }\textbf {\bibinfo {volume} {326}},\ \bibinfo {pages}
  {96} (\bibinfo {year} {2011})}\BibitemShut {NoStop}%
\bibitem [{\citenamefont {Saffman}\ \emph {et~al.}(2010)\citenamefont
  {Saffman}, \citenamefont {Walker},\ and\ \citenamefont
  {M\o{}lmer}}]{Saffman2010}%
  \BibitemOpen
  \bibfield  {author} {\bibinfo {author} {\bibfnamefont {M.}~\bibnamefont
  {Saffman}}, \bibinfo {author} {\bibfnamefont {T.~G.}\ \bibnamefont
  {Walker}},\ and\ \bibinfo {author} {\bibfnamefont {K.}~\bibnamefont
  {M\o{}lmer}},\ }\bibfield  {title} {\bibinfo {title} {Quantum information
  with rydberg atoms},\ }\href {https://doi.org/10.1103/RevModPhys.82.2313}
  {\bibfield  {journal} {\bibinfo  {journal} {Rev. Mod. Phys.}\ }\textbf
  {\bibinfo {volume} {82}},\ \bibinfo {pages} {2313} (\bibinfo {year}
  {2010})}\BibitemShut {NoStop}%
\bibitem [{\citenamefont {Henriet}\ \emph {et~al.}(2020)\citenamefont
  {Henriet}, \citenamefont {Beguin}, \citenamefont {Signoles}, \citenamefont
  {Lahaye}, \citenamefont {Browaeys}, \citenamefont {Reymond},\ and\
  \citenamefont {Jurczak}}]{henriet_quantum_2020}%
  \BibitemOpen
  \bibfield  {author} {\bibinfo {author} {\bibfnamefont {L.}~\bibnamefont
  {Henriet}}, \bibinfo {author} {\bibfnamefont {L.}~\bibnamefont {Beguin}},
  \bibinfo {author} {\bibfnamefont {A.}~\bibnamefont {Signoles}}, \bibinfo
  {author} {\bibfnamefont {T.}~\bibnamefont {Lahaye}}, \bibinfo {author}
  {\bibfnamefont {A.}~\bibnamefont {Browaeys}}, \bibinfo {author}
  {\bibfnamefont {G.-O.}\ \bibnamefont {Reymond}},\ and\ \bibinfo {author}
  {\bibfnamefont {C.}~\bibnamefont {Jurczak}},\ }\bibfield  {title} {\bibinfo
  {title} {Quantum computing with neutral atoms},\ }\href
  {https://doi.org/10.22331/q-2020-09-21-327} {\bibfield  {journal} {\bibinfo
  {journal} {Quantum}\ }\textbf {\bibinfo {volume} {4}},\ \bibinfo {pages}
  {327} (\bibinfo {year} {2020})}\BibitemShut {NoStop}%
\bibitem [{\citenamefont {Browaeys}\ and\ \citenamefont
  {Lahaye}(2020)}]{browaeys_many-body_2020}%
  \BibitemOpen
  \bibfield  {author} {\bibinfo {author} {\bibfnamefont {A.}~\bibnamefont
  {Browaeys}}\ and\ \bibinfo {author} {\bibfnamefont {T.}~\bibnamefont
  {Lahaye}},\ }\bibfield  {title} {\bibinfo {title} {Many-body physics with
  individually controlled {R}ydberg atoms},\ }\href
  {https://doi.org/10.1038/s41567-019-0733-z} {\bibfield  {journal} {\bibinfo
  {journal} {Nat. Phys.}\ }\textbf {\bibinfo {volume} {16}},\ \bibinfo {pages}
  {132} (\bibinfo {year} {2020})}\BibitemShut {NoStop}%
\bibitem [{\citenamefont
  {de~L{\ifmmode\acute{e}\else\'{e}\fi}s{\ifmmode\acute{e}\else\'{e}\fi}leuc}\
  \emph {et~al.}(2019)\citenamefont
  {de~L{\ifmmode\acute{e}\else\'{e}\fi}s{\ifmmode\acute{e}\else\'{e}\fi}leuc},
  \citenamefont {Lienhard}, \citenamefont {Scholl}, \citenamefont {Barredo},
  \citenamefont {Weber}, \citenamefont {Lang}, \citenamefont
  {B{\ifmmode\ddot{u}\else\"{u}\fi}chler}, \citenamefont {Lahaye},\ and\
  \citenamefont {Browaeys}}]{deLeseleuc2019}%
  \BibitemOpen
  \bibfield  {author} {\bibinfo {author} {\bibfnamefont {S.}~\bibnamefont
  {de~L{\ifmmode\acute{e}\else\'{e}\fi}s{\ifmmode\acute{e}\else\'{e}\fi}leuc}},
  \bibinfo {author} {\bibfnamefont {V.}~\bibnamefont {Lienhard}}, \bibinfo
  {author} {\bibfnamefont {P.}~\bibnamefont {Scholl}}, \bibinfo {author}
  {\bibfnamefont {D.}~\bibnamefont {Barredo}}, \bibinfo {author} {\bibfnamefont
  {S.}~\bibnamefont {Weber}}, \bibinfo {author} {\bibfnamefont
  {N.}~\bibnamefont {Lang}}, \bibinfo {author} {\bibfnamefont {H.~P.}\
  \bibnamefont {B{\ifmmode\ddot{u}\else\"{u}\fi}chler}}, \bibinfo {author}
  {\bibfnamefont {T.}~\bibnamefont {Lahaye}},\ and\ \bibinfo {author}
  {\bibfnamefont {A.}~\bibnamefont {Browaeys}},\ }\bibfield  {title} {\bibinfo
  {title} {{Observation of a symmetry-protected topological phase of
  interacting bosons with Rydberg atoms}},\ }\href
  {https://www.science.org/doi/10.1126/science.aav9105} {\bibfield  {journal}
  {\bibinfo  {journal} {Science}\ }\textbf {\bibinfo {volume} {365}},\ \bibinfo
  {pages} {775} (\bibinfo {year} {2019})}\BibitemShut {NoStop}%
\bibitem [{\citenamefont {Scholl}\ \emph {et~al.}(2022)\citenamefont {Scholl},
  \citenamefont {Williams}, \citenamefont {Bornet}, \citenamefont {Wallner},
  \citenamefont {Barredo}, \citenamefont {Henriet}, \citenamefont {Signoles},
  \citenamefont {Hainaut}, \citenamefont {Franz}, \citenamefont {Geier},
  \citenamefont {Tebben}, \citenamefont {Salzinger}, \citenamefont {Z\"urn},
  \citenamefont {Lahaye}, \citenamefont {Weidem\"uller},\ and\ \citenamefont
  {Browaeys}}]{Scholl2022}%
  \BibitemOpen
  \bibfield  {author} {\bibinfo {author} {\bibfnamefont {P.}~\bibnamefont
  {Scholl}}, \bibinfo {author} {\bibfnamefont {H.~J.}\ \bibnamefont
  {Williams}}, \bibinfo {author} {\bibfnamefont {G.}~\bibnamefont {Bornet}},
  \bibinfo {author} {\bibfnamefont {F.}~\bibnamefont {Wallner}}, \bibinfo
  {author} {\bibfnamefont {D.}~\bibnamefont {Barredo}}, \bibinfo {author}
  {\bibfnamefont {L.}~\bibnamefont {Henriet}}, \bibinfo {author} {\bibfnamefont
  {A.}~\bibnamefont {Signoles}}, \bibinfo {author} {\bibfnamefont
  {C.}~\bibnamefont {Hainaut}}, \bibinfo {author} {\bibfnamefont
  {T.}~\bibnamefont {Franz}}, \bibinfo {author} {\bibfnamefont
  {S.}~\bibnamefont {Geier}}, \bibinfo {author} {\bibfnamefont
  {A.}~\bibnamefont {Tebben}}, \bibinfo {author} {\bibfnamefont
  {A.}~\bibnamefont {Salzinger}}, \bibinfo {author} {\bibfnamefont
  {G.}~\bibnamefont {Z\"urn}}, \bibinfo {author} {\bibfnamefont
  {T.}~\bibnamefont {Lahaye}}, \bibinfo {author} {\bibfnamefont
  {M.}~\bibnamefont {Weidem\"uller}},\ and\ \bibinfo {author} {\bibfnamefont
  {A.}~\bibnamefont {Browaeys}},\ }\bibfield  {title} {\bibinfo {title}
  {Microwave engineering of programmable $xxz$ hamiltonians in arrays of
  rydberg atoms},\ }\href {https://doi.org/10.1103/PRXQuantum.3.020303}
  {\bibfield  {journal} {\bibinfo  {journal} {PRX Quantum}\ }\textbf {\bibinfo
  {volume} {3}},\ \bibinfo {pages} {020303} (\bibinfo {year}
  {2022})}\BibitemShut {NoStop}%
\bibitem [{\citenamefont {Hatsugai}\ and\ \citenamefont
  {Fukui}(2016)}]{Hatsugai_bulkedge_2016}%
  \BibitemOpen
  \bibfield  {author} {\bibinfo {author} {\bibfnamefont {Y.}~\bibnamefont
  {Hatsugai}}\ and\ \bibinfo {author} {\bibfnamefont {T.}~\bibnamefont
  {Fukui}},\ }\bibfield  {title} {\bibinfo {title} {Bulk-edge correspondence in
  topological pumping},\ }\href {https://doi.org/10.1103/PhysRevB.94.041102}
  {\bibfield  {journal} {\bibinfo  {journal} {Phys. Rev. B}\ }\textbf {\bibinfo
  {volume} {94}},\ \bibinfo {pages} {041102} (\bibinfo {year}
  {2016})}\BibitemShut {NoStop}%
\bibitem [{\citenamefont {Nogrette}\ \emph {et~al.}(2014)\citenamefont
  {Nogrette}, \citenamefont {Labuhn}, \citenamefont {Ravets}, \citenamefont
  {Barredo}, \citenamefont {Béguin}, \citenamefont {Vernier}, \citenamefont
  {Lahaye},\ and\ \citenamefont {Browaeys}}]{Nogrette2014}%
  \BibitemOpen
  \bibfield  {author} {\bibinfo {author} {\bibfnamefont {F.}~\bibnamefont
  {Nogrette}}, \bibinfo {author} {\bibfnamefont {H.}~\bibnamefont {Labuhn}},
  \bibinfo {author} {\bibfnamefont {S.}~\bibnamefont {Ravets}}, \bibinfo
  {author} {\bibfnamefont {D.}~\bibnamefont {Barredo}}, \bibinfo {author}
  {\bibfnamefont {L.}~\bibnamefont {Béguin}}, \bibinfo {author} {\bibfnamefont
  {A.}~\bibnamefont {Vernier}}, \bibinfo {author} {\bibfnamefont
  {T.}~\bibnamefont {Lahaye}},\ and\ \bibinfo {author} {\bibfnamefont
  {A.}~\bibnamefont {Browaeys}},\ }\bibfield  {title} {\bibinfo {title}
  {Single-atom trapping in holographic {2D} arrays of microtraps with arbitrary
  geometries},\ }\href {https://doi.org/10.1103/PhysRevX.4.021034} {\bibfield
  {journal} {\bibinfo  {journal} {Phys. Rev. X}\ }\textbf {\bibinfo {volume}
  {4}},\ \bibinfo {pages} {021034} (\bibinfo {year} {2014})}\BibitemShut
  {NoStop}%
\bibitem [{\citenamefont {Ebadi}\ \emph {et~al.}(2022)\citenamefont {Ebadi},
  \citenamefont {Keesling}, \citenamefont {Cain}, \citenamefont {Wang},
  \citenamefont {Levine}, \citenamefont {Bluvstein}, \citenamefont {Semeghini},
  \citenamefont {Omran}, \citenamefont {Liu}, \citenamefont {Samajdar},
  \citenamefont {Luo}, \citenamefont {Nash}, \citenamefont {Gao}, \citenamefont
  {Barak}, \citenamefont {Farhi}, \citenamefont {Sachdev}, \citenamefont
  {Gemelke}, \citenamefont {Zhou}, \citenamefont {Choi}, \citenamefont
  {Pichler}, \citenamefont {Wang}, \citenamefont {Greiner}, \citenamefont
  {Vuleti{\'c}},\ and\ \citenamefont {Lukin}}]{ebadiQuantum2022}%
  \BibitemOpen
  \bibfield  {author} {\bibinfo {author} {\bibfnamefont {S.}~\bibnamefont
  {Ebadi}}, \bibinfo {author} {\bibfnamefont {A.}~\bibnamefont {Keesling}},
  \bibinfo {author} {\bibfnamefont {M.}~\bibnamefont {Cain}}, \bibinfo {author}
  {\bibfnamefont {T.~T.}\ \bibnamefont {Wang}}, \bibinfo {author}
  {\bibfnamefont {H.}~\bibnamefont {Levine}}, \bibinfo {author} {\bibfnamefont
  {D.}~\bibnamefont {Bluvstein}}, \bibinfo {author} {\bibfnamefont
  {G.}~\bibnamefont {Semeghini}}, \bibinfo {author} {\bibfnamefont
  {A.}~\bibnamefont {Omran}}, \bibinfo {author} {\bibfnamefont {J.-G.}\
  \bibnamefont {Liu}}, \bibinfo {author} {\bibfnamefont {R.}~\bibnamefont
  {Samajdar}}, \bibinfo {author} {\bibfnamefont {X.-Z.}\ \bibnamefont {Luo}},
  \bibinfo {author} {\bibfnamefont {B.}~\bibnamefont {Nash}}, \bibinfo {author}
  {\bibfnamefont {X.}~\bibnamefont {Gao}}, \bibinfo {author} {\bibfnamefont
  {B.}~\bibnamefont {Barak}}, \bibinfo {author} {\bibfnamefont
  {E.}~\bibnamefont {Farhi}}, \bibinfo {author} {\bibfnamefont
  {S.}~\bibnamefont {Sachdev}}, \bibinfo {author} {\bibfnamefont
  {N.}~\bibnamefont {Gemelke}}, \bibinfo {author} {\bibfnamefont
  {L.}~\bibnamefont {Zhou}}, \bibinfo {author} {\bibfnamefont {S.}~\bibnamefont
  {Choi}}, \bibinfo {author} {\bibfnamefont {H.}~\bibnamefont {Pichler}},
  \bibinfo {author} {\bibfnamefont {S.-T.}\ \bibnamefont {Wang}}, \bibinfo
  {author} {\bibfnamefont {M.}~\bibnamefont {Greiner}}, \bibinfo {author}
  {\bibfnamefont {V.}~\bibnamefont {Vuleti{\'c}}},\ and\ \bibinfo {author}
  {\bibfnamefont {M.~D.}\ \bibnamefont {Lukin}},\ }\bibfield  {title} {\bibinfo
  {title} {Quantum optimization of maximum independent set using rydberg atom
  arrays},\ }\href {https://doi.org/10.1126/science.abo6587} {\bibfield
  {journal} {\bibinfo  {journal} {Science}\ }\textbf {\bibinfo {volume}
  {376}},\ \bibinfo {pages} {1209} (\bibinfo {year} {2022})}\BibitemShut
  {NoStop}%
\bibitem [{\citenamefont {Barredo}\ \emph {et~al.}(2018)\citenamefont
  {Barredo}, \citenamefont {Lienhard}, \citenamefont {{de L{\'e}s{\'e}leuc}},
  \citenamefont {Lahaye},\ and\ \citenamefont
  {Browaeys}}]{barredoSynthetic2018}%
  \BibitemOpen
  \bibfield  {author} {\bibinfo {author} {\bibfnamefont {D.}~\bibnamefont
  {Barredo}}, \bibinfo {author} {\bibfnamefont {V.}~\bibnamefont {Lienhard}},
  \bibinfo {author} {\bibfnamefont {S.}~\bibnamefont {{de L{\'e}s{\'e}leuc}}},
  \bibinfo {author} {\bibfnamefont {T.}~\bibnamefont {Lahaye}},\ and\ \bibinfo
  {author} {\bibfnamefont {A.}~\bibnamefont {Browaeys}},\ }\bibfield  {title}
  {\bibinfo {title} {Synthetic three-dimensional atomic structures assembled
  atom by atom},\ }\href {https://doi.org/10.1038/s41586-018-0450-2} {\bibfield
   {journal} {\bibinfo  {journal} {Nature}\ }\textbf {\bibinfo {volume}
  {561}},\ \bibinfo {pages} {79} (\bibinfo {year} {2018})}\BibitemShut
  {NoStop}%
\bibitem [{\citenamefont {Chen}\ \emph {et~al.}(2023)\citenamefont {Chen},
  \citenamefont {Bornet}, \citenamefont {Bintz}, \citenamefont {Emperauger},
  \citenamefont {Leclerc}, \citenamefont {Liu}, \citenamefont {Scholl},
  \citenamefont {Barredo}, \citenamefont {Hauschild}, \citenamefont
  {Chatterjee}, \citenamefont {Schuler}, \citenamefont {L{\"a}uchli},
  \citenamefont {Zaletel}, \citenamefont {Lahaye}, \citenamefont {Yao},\ and\
  \citenamefont {Browaeys}}]{chenContinuous2023}%
  \BibitemOpen
  \bibfield  {author} {\bibinfo {author} {\bibfnamefont {C.}~\bibnamefont
  {Chen}}, \bibinfo {author} {\bibfnamefont {G.}~\bibnamefont {Bornet}},
  \bibinfo {author} {\bibfnamefont {M.}~\bibnamefont {Bintz}}, \bibinfo
  {author} {\bibfnamefont {G.}~\bibnamefont {Emperauger}}, \bibinfo {author}
  {\bibfnamefont {L.}~\bibnamefont {Leclerc}}, \bibinfo {author} {\bibfnamefont
  {V.~S.}\ \bibnamefont {Liu}}, \bibinfo {author} {\bibfnamefont
  {P.}~\bibnamefont {Scholl}}, \bibinfo {author} {\bibfnamefont
  {D.}~\bibnamefont {Barredo}}, \bibinfo {author} {\bibfnamefont
  {J.}~\bibnamefont {Hauschild}}, \bibinfo {author} {\bibfnamefont
  {S.}~\bibnamefont {Chatterjee}}, \bibinfo {author} {\bibfnamefont
  {M.}~\bibnamefont {Schuler}}, \bibinfo {author} {\bibfnamefont {A.~M.}\
  \bibnamefont {L{\"a}uchli}}, \bibinfo {author} {\bibfnamefont {M.~P.}\
  \bibnamefont {Zaletel}}, \bibinfo {author} {\bibfnamefont {T.}~\bibnamefont
  {Lahaye}}, \bibinfo {author} {\bibfnamefont {N.~Y.}\ \bibnamefont {Yao}},\
  and\ \bibinfo {author} {\bibfnamefont {A.}~\bibnamefont {Browaeys}},\
  }\bibfield  {title} {\bibinfo {title} {Continuous symmetry breaking in a
  two-dimensional {{Rydberg}} array},\ }\href
  {https://doi.org/10.1038/s41586-023-05859-2} {\bibfield  {journal} {\bibinfo
  {journal} {Nature}\ }\textbf {\bibinfo {volume} {616}},\ \bibinfo {pages}
  {691} (\bibinfo {year} {2023})}\BibitemShut {NoStop}%
\bibitem [{\citenamefont {Geier}\ \emph {et~al.}(2021)\citenamefont {Geier},
  \citenamefont {Thaicharoen}, \citenamefont {Hainaut}, \citenamefont {Franz},
  \citenamefont {Salzinger}, \citenamefont {Tebben}, \citenamefont
  {Grimshandl}, \citenamefont {Zürn},\ and\ \citenamefont
  {Weidemüller}}]{geierFloquet2021b}%
  \BibitemOpen
  \bibfield  {author} {\bibinfo {author} {\bibfnamefont {S.}~\bibnamefont
  {Geier}}, \bibinfo {author} {\bibfnamefont {N.}~\bibnamefont {Thaicharoen}},
  \bibinfo {author} {\bibfnamefont {C.}~\bibnamefont {Hainaut}}, \bibinfo
  {author} {\bibfnamefont {T.}~\bibnamefont {Franz}}, \bibinfo {author}
  {\bibfnamefont {A.}~\bibnamefont {Salzinger}}, \bibinfo {author}
  {\bibfnamefont {A.}~\bibnamefont {Tebben}}, \bibinfo {author} {\bibfnamefont
  {D.}~\bibnamefont {Grimshandl}}, \bibinfo {author} {\bibfnamefont
  {G.}~\bibnamefont {Zürn}},\ and\ \bibinfo {author} {\bibfnamefont
  {M.}~\bibnamefont {Weidemüller}},\ }\bibfield  {title} {\bibinfo {title}
  {Floquet {Hamiltonian} engineering of an isolated many-body spin system},\
  }\href {https://doi.org/10.1126/science.abd9547} {\bibfield  {journal}
  {\bibinfo  {journal} {Science}\ }\textbf {\bibinfo {volume} {374}},\ \bibinfo
  {pages} {1149} (\bibinfo {year} {2021})}\BibitemShut {NoStop}%
\bibitem [{\citenamefont {Vandersypen}\ and\ \citenamefont
  {Chuang}(2005)}]{vandersypenNMR2005a}%
  \BibitemOpen
  \bibfield  {author} {\bibinfo {author} {\bibfnamefont {L.~M.~K.}\
  \bibnamefont {Vandersypen}}\ and\ \bibinfo {author} {\bibfnamefont {I.~L.}\
  \bibnamefont {Chuang}},\ }\bibfield  {title} {\bibinfo {title} {{NMR}
  techniques for quantum control and computation},\ }\href
  {https://doi.org/10.1103/RevModPhys.76.1037} {\bibfield  {journal} {\bibinfo
  {journal} {Rev. Mod. Phys.}\ }\textbf {\bibinfo {volume} {76}},\ \bibinfo
  {pages} {1037} (\bibinfo {year} {2005})}\BibitemShut {NoStop}%
\bibitem [{\citenamefont {Weckesser}\ \emph {et~al.}(2024)\citenamefont
  {Weckesser}, \citenamefont {Srakaew}, \citenamefont {Blatz}, \citenamefont
  {Wei}, \citenamefont {Adler}, \citenamefont {Agrawal}, \citenamefont
  {Bohrdt}, \citenamefont {Bloch},\ and\ \citenamefont
  {Zeiher}}]{weckesser2024realization}%
  \BibitemOpen
  \bibfield  {author} {\bibinfo {author} {\bibfnamefont {P.}~\bibnamefont
  {Weckesser}}, \bibinfo {author} {\bibfnamefont {K.}~\bibnamefont {Srakaew}},
  \bibinfo {author} {\bibfnamefont {T.}~\bibnamefont {Blatz}}, \bibinfo
  {author} {\bibfnamefont {D.}~\bibnamefont {Wei}}, \bibinfo {author}
  {\bibfnamefont {D.}~\bibnamefont {Adler}}, \bibinfo {author} {\bibfnamefont
  {S.}~\bibnamefont {Agrawal}}, \bibinfo {author} {\bibfnamefont
  {A.}~\bibnamefont {Bohrdt}}, \bibinfo {author} {\bibfnamefont
  {I.}~\bibnamefont {Bloch}},\ and\ \bibinfo {author} {\bibfnamefont
  {J.}~\bibnamefont {Zeiher}},\ }\href@noop {} {\bibinfo {title} {Realization
  of a rydberg-dressed extended bose hubbard model}} (\bibinfo {year} {2024}),\
  \Eprint {https://arxiv.org/abs/2405.20128} {arXiv:2405.20128
  [cond-mat.quant-gas]} \BibitemShut {NoStop}%
\bibitem [{\citenamefont {Marks}\ \emph {et~al.}(2021)\citenamefont {Marks},
  \citenamefont {Sch\"uler}, \citenamefont {Budich},\ and\ \citenamefont
  {Devereaux}}]{Marks2021}%
  \BibitemOpen
  \bibfield  {author} {\bibinfo {author} {\bibfnamefont {J.~A.}\ \bibnamefont
  {Marks}}, \bibinfo {author} {\bibfnamefont {M.}~\bibnamefont {Sch\"uler}},
  \bibinfo {author} {\bibfnamefont {J.~C.}\ \bibnamefont {Budich}},\ and\
  \bibinfo {author} {\bibfnamefont {T.~P.}\ \bibnamefont {Devereaux}},\
  }\bibfield  {title} {\bibinfo {title} {Correlation-assisted quantized charge
  pumping},\ }\href {https://doi.org/10.1103/PhysRevB.103.035112} {\bibfield
  {journal} {\bibinfo  {journal} {Phys. Rev. B}\ }\textbf {\bibinfo {volume}
  {103}},\ \bibinfo {pages} {035112} (\bibinfo {year} {2021})}\BibitemShut
  {NoStop}%
\bibitem [{\citenamefont {Cornish}\ \emph {et~al.}(2024)\citenamefont
  {Cornish}, \citenamefont {Tarbutt},\ and\ \citenamefont
  {Hazzard}}]{cornish2024quantum}%
  \BibitemOpen
  \bibfield  {author} {\bibinfo {author} {\bibfnamefont {S.~L.}\ \bibnamefont
  {Cornish}}, \bibinfo {author} {\bibfnamefont {M.~R.}\ \bibnamefont
  {Tarbutt}},\ and\ \bibinfo {author} {\bibfnamefont {K.~R.~A.}\ \bibnamefont
  {Hazzard}},\ }\href@noop {} {\bibinfo {title} {Quantum computation and
  quantum simulation with ultracold molecules}} (\bibinfo {year} {2024}),\
  \Eprint {https://arxiv.org/abs/2401.05086} {arXiv:2401.05086
  [cond-mat.quant-gas]} \BibitemShut {NoStop}%
\bibitem [{\citenamefont {M\'arquez}\ \emph {et~al.}(2024)\citenamefont
  {M\'arquez}, \citenamefont {Aucar~Boidi}, \citenamefont {Hallberg},\ and\
  \citenamefont {Aligia}}]{PhysRevB.109.235143}%
  \BibitemOpen
  \bibfield  {author} {\bibinfo {author} {\bibfnamefont {B.~F.}\ \bibnamefont
  {M\'arquez}}, \bibinfo {author} {\bibfnamefont {N.}~\bibnamefont
  {Aucar~Boidi}}, \bibinfo {author} {\bibfnamefont {K.}~\bibnamefont
  {Hallberg}},\ and\ \bibinfo {author} {\bibfnamefont {A.~A.}\ \bibnamefont
  {Aligia}},\ }\bibfield  {title} {\bibinfo {title} {Phase diagram and topology
  of the xxz chain with alternating bonds and staggered magnetic field},\
  }\href {https://doi.org/10.1103/PhysRevB.109.235143} {\bibfield  {journal}
  {\bibinfo  {journal} {Phys. Rev. B}\ }\textbf {\bibinfo {volume} {109}},\
  \bibinfo {pages} {235143} (\bibinfo {year} {2024})}\BibitemShut {NoStop}%
\bibitem [{\citenamefont {Hauschild}\ and\ \citenamefont
  {Pollmann}(2018)}]{tenpy}%
  \BibitemOpen
  \bibfield  {author} {\bibinfo {author} {\bibfnamefont {J.}~\bibnamefont
  {Hauschild}}\ and\ \bibinfo {author} {\bibfnamefont {F.}~\bibnamefont
  {Pollmann}},\ }\bibfield  {title} {\bibinfo {title} {{Efficient numerical
  simulations with Tensor Networks: Tensor Network Python (TeNPy)}},\ }\href
  {https://doi.org/10.21468/SciPostPhysLectNotes.5} {\bibfield  {journal}
  {\bibinfo  {journal} {SciPost Phys. Lect. Notes}\ ,\ \bibinfo {pages} {5}}
  (\bibinfo {year} {2018})}\BibitemShut {NoStop}%
\bibitem [{\citenamefont {Zaletel}\ \emph {et~al.}(2014)\citenamefont
  {Zaletel}, \citenamefont {Mong},\ and\ \citenamefont
  {Pollmann}}]{Zaletel2014}%
  \BibitemOpen
  \bibfield  {author} {\bibinfo {author} {\bibfnamefont {M.~P.}\ \bibnamefont
  {Zaletel}}, \bibinfo {author} {\bibfnamefont {R.~S.~K.}\ \bibnamefont
  {Mong}},\ and\ \bibinfo {author} {\bibfnamefont {F.}~\bibnamefont
  {Pollmann}},\ }\bibfield  {title} {\bibinfo {title} {Flux insertion,
  entanglement, and quantized responses},\ }\href
  {https://doi.org/10.1088/1742-5468/2014/10/p10007} {\bibfield  {journal}
  {\bibinfo  {journal} {J. Stat. Mech. Theory Exp.}\ }\textbf {\bibinfo
  {volume} {2014}},\ \bibinfo {pages} {P10007} (\bibinfo {year}
  {2014})}\BibitemShut {NoStop}%
\end{thebibliography}
\end{document}